\documentclass[amssymb,amsmath, aps,floats,preprint,prb]{revtex4}
\usepackage{color,graphicx,pstricks,subfigure}
\usepackage{bm}
\usepackage[dvips]{rotating}
\usepackage{epstopdf}
\usepackage[abs]{overpic}
\usepackage{appendix}
\makeatletter\renewcommand\@biblabel[1]{{#1}}\makeatother

\def\k{{\mathbf{k}}}
\def\R{{\mathbf{R}}}
\def\q{{\mathbf{q}}}

\begin{document}

\title{\,
%\vspace*{4 mm}
%$\mbox{
Disorder- and Field-Induced Antiferromagnetism in \newline Cuprate
Superconductors
%}$
%\vspace{2cm}}
}
\author{{}
Markus~Schmid,$^{1*}$ Brian~M.~Andersen,$^2$ Arno~P.~Kampf,$^1$ and
P.~J.~Hirschfeld$^3$ \vspace{0,5cm}}

\affiliation{$^1$Theoretical Physics III, Center for Electronic Correlations and Magnetism, Institute
of Physics, University of Augsburg,
D-86135 Augsburg, Germany
\\
$^2$Niels Bohr Institute, University of Copenhagen, Universitetsparken 5,
DK-2100 Copenhagen, Denmark
\\
$^3$Department of Physics, University of Florida, Gainesville, FL 32611, USA\vspace{1cm}}

\date{\today}

\begin{abstract}
{\bf \boldmath The underdoped high-$T_c$ materials are characterized
by a competition between Cooper pairing and antiferromagnetic (AF) order. Important
differences between the superconducting (SC) state of these materials and
conventional superconductors include the $d$-wave pairing symmetry and
a remarkable magnetic response to nonmagnetic perturbations, whereby droplets of
spin-density wave (SDW) order can form around impurities and the cores of vortices.
In a simple picture, whenever SC is suppressed locally,
SDW order is nucleated. Within a mean-field theory of $d$-wave SC in an applied
magnetic field including disorder and Hubbard correlations, we show in fact
that the creation of SDW order is not simply due to suppression of the
SC order parameter, but rather due to a correlation-induced
splitting of the electronic bound state created by the perturbation. 
Since the bound state exists because of the sign change of the order
parameter along quasiparticle trajectories, the induced SDW order is a direct
consequence of the $d$-wave symmetry. 
%We further show the existence of a novel
%interplay between the magnetic order induced by disorder and applied field
%respectively, and find that 
Furthermore the formation of anti-phase domain walls is important for obtaining the 
correct temperature dependence of the induced magnetism as measured by neutron diffraction.}
\\
\\
$^*$ Corresponding author e-mail: markus.schmid@physik.uni-augsburg.de

 %We explore the field
% and the temperature dependence of the induced magnetization in the presence of disorder and magnetic field alone, and
%subsequently show that the two phenomena interfere and may not be treated independently. The theory is capable to describe
%a transition
% from a weakly disordered state of isolated magnetic droplets to a state with disordered stripe patterns.
%Many of the puzzling experimental
% data pertaining to the inherent ``spin-glass" phases, which
% persist into the superconducting dome of intrinsically disordered cuprates, are hereby elucidated. }
\end{abstract}

%\vspace{1cm}

\maketitle

%%%%%%%%%%

A superconductor is characterized by a
Bardeen-Cooper-Schrieffer (BCS) order parameter $\Delta_\k(\R)$, where
$\R$ is the center-of-mass coordinate of a Cooper pair of electrons
with momenta $(\k,-\k)$. The bulk ground state of such a system is
homogeneous, but a spatial perturbation which breaks pairs, e.g. a
magnetic impurity, may  cause the suppression of $\Delta_\k(\R)$
locally.  What is revealed when SC is suppressed is
the electronic phase in the absence of $\Delta_\k$, a normal Fermi
liquid.  Thus the low-energy excitations near magnetic impurities
and in the vortex cores of conventional SC are
essentially Landau quasiparticles trapped in bound states. The
 underdoped cuprates have been studied intensively in recent years in part
because their proximity to the Mott insulator is thought to be
responsible for many unusual properties, including possibly
high-temperature SC itself.  These systems are
quite different from conventional SC, because when
the pair amplitude is suppressed locally, e.g. by a vortex, a
competing ordered state\cite{SachdevRMP} stabilized by the
proximity to the Mott state appears to emerge instead of a normal
metal.
% In the intrinsically  disordered cuprates like
%La$_{2-x}$Sr$_x$CuO$_4$ (LSCO) and Bi$_2$Sr$_2$CaCu$_2$O$_{8+\delta}$
%(BSCCO), 
This state is characterized at low temperatures $T$ by static local
SDW order with an ordering wave vector $\bf Q$ near
$(\pi,\pi)$, an order which is not present in the state above $T_c$. This was
reported first in elastic neutron scattering experiments\cite{Lake} on La$_{2-x}$Sr$_x$CuO$_4$  
(LSCO), with a correlation length
of several hundred \AA, 
but has been confirmed in other underdoped cuprate materials as well.\cite{khaykovich02,khaykovich05,chang08,Haug} An enhancement of
incommensurate static order was observed with increasing the applied
magnetic field up to 14T.\cite{Lake} Because the signal disappeared above
$T_c$, the magnetism was attributed to the vortices; 
%in Ref. \onlinecite{Lake}
%an estimate for the AF correlation length of $\sim 400$\AA was
%extracted. 
indeed, scanning tunnelling microscopy (STM) measurements\cite{Hoffman}
on Bi$_2$Sr$_2$CaCu$_2$O$_{8+\delta}$ (BSSCO)  were able to directly image unusual charge order near the vortex
cores which is almost certainly related to the field-induced SDW detected by
neutron scattering.

\begin{figure}[t!]
\hspace{.1cm}
\begin{overpic}[scale=.55,unit=1mm]{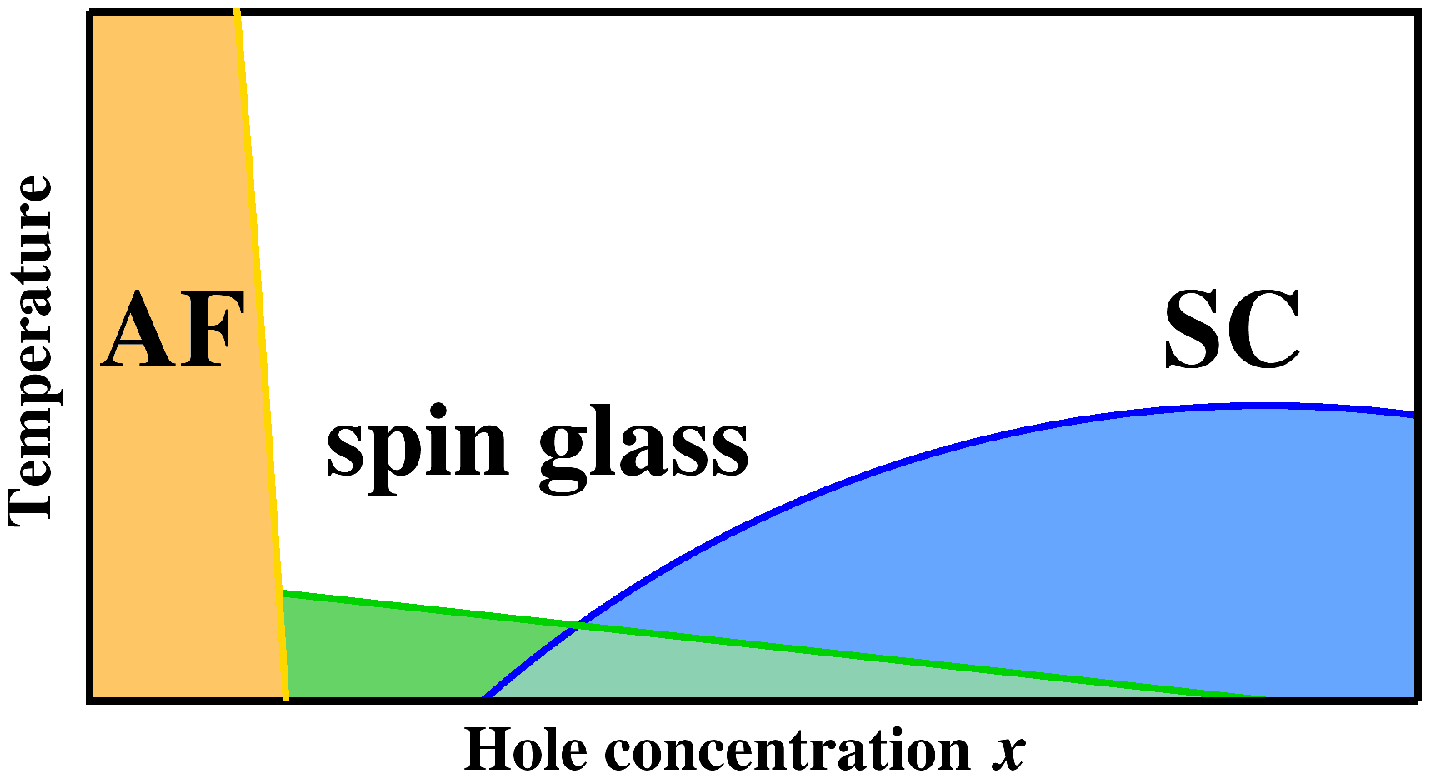}
% \begin{overpic}[scale=.75,unit=1mm]{fig0a.eps}
    \put(-3,42){{\bf a}}
 \end{overpic}
\hspace{.6cm}
\begin{overpic}[scale=.26,unit=1mm]{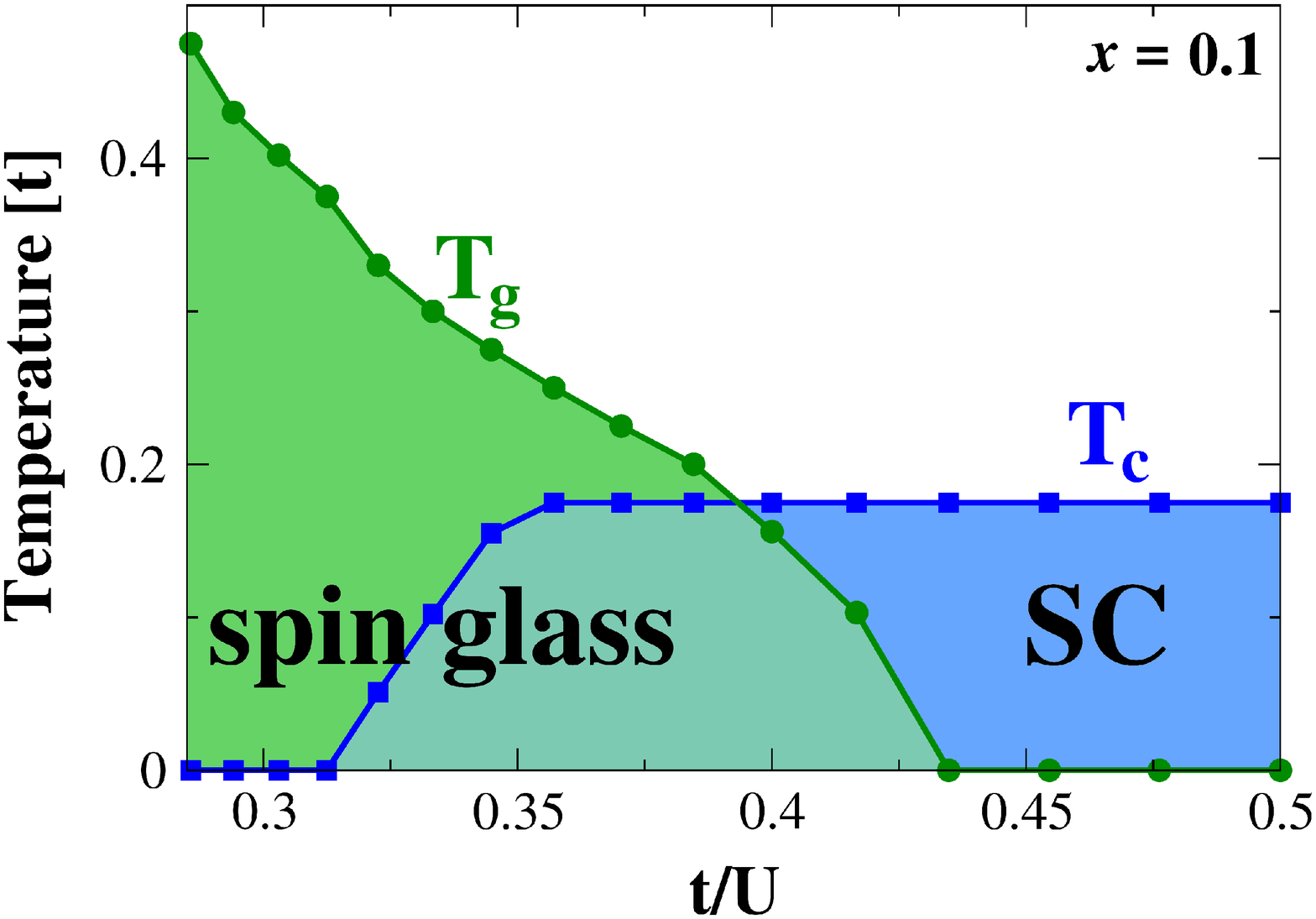}
    \put(-3,42){{\bf b}}
 \end{overpic}

\vspace{1em}
\hspace{.1cm}
 \begin{overpic}[scale=.265,unit=1mm]{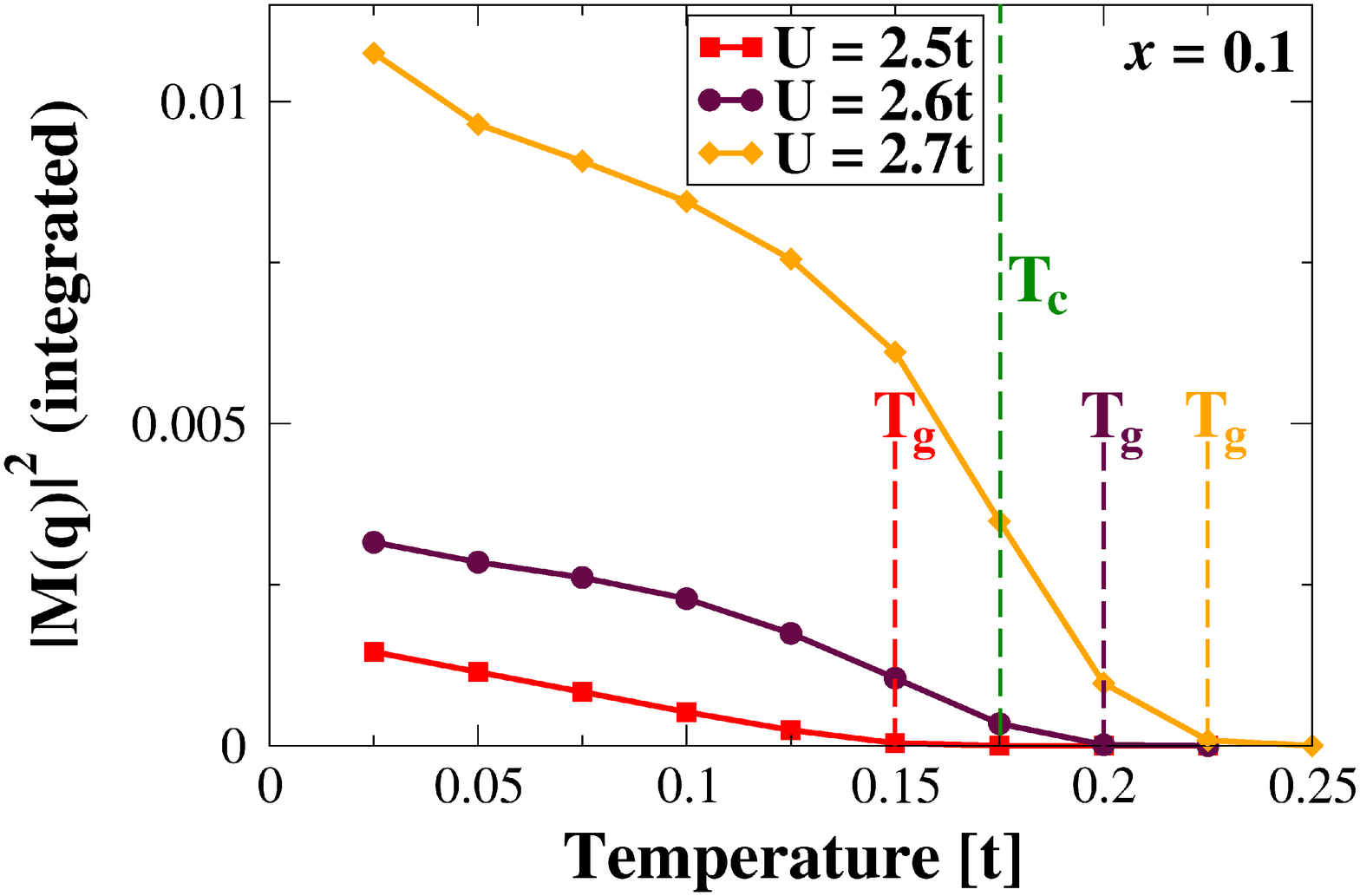}
    \put(-3,47){{\bf c}}
 \end{overpic}
\hspace{1.3cm}
 \begin{overpic}[scale=.265,unit=1mm]{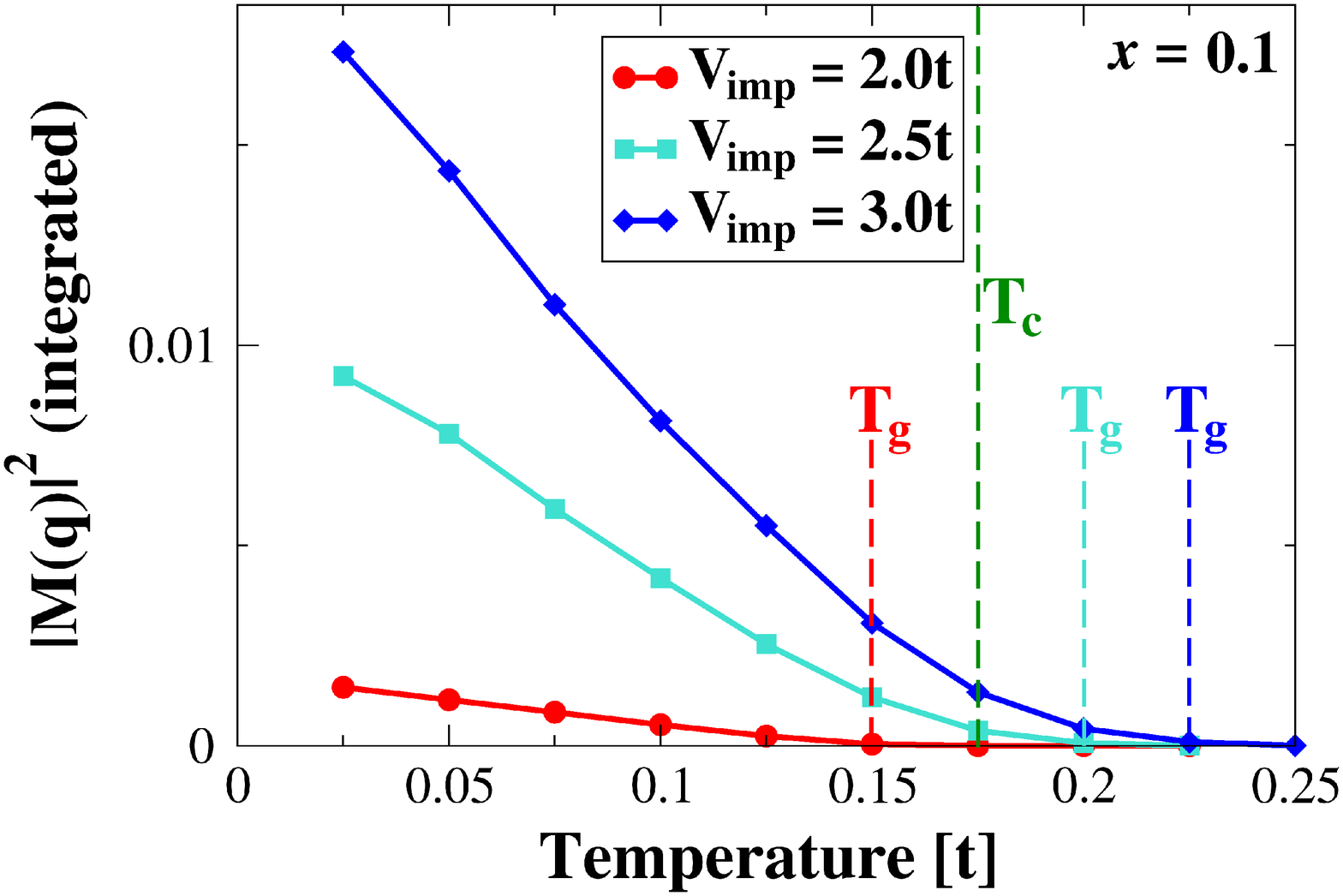}
    \put(-3,47){{\bf d}}
 \end{overpic}
\caption{\setlength{\baselineskip}{5mm} {\bf \boldmath Phase
diagrams and temperature dependence of magnetic order.} {\bf a}.
Schematic temperature $T$ vs. doping $x$ phase diagram for cuprates. {\bf
b}. $t/U$ dependence of the SC $T_c$ and the
disorder-induced magnetic transition temperature $T_g$ for an
impurity concentration $n_{imp} = 10\%$ with potential strength
$V_{imp} = 2.0t$ and a hole concentration $x=0.1$. $t$ is the
nearest-neighbor hopping amplitude and $U$ denotes the Hubbard on-site
repulsion. {\bf c, d}. Magnetic Fourier component $|M({\bf q})|^2$ at the
ordering wavevector integrated around ($\pi, \pi$) vs. temperature for
({\bf c}) different interaction strengths $U$ (at fixed $V_{imp} = 2.0t$)
and ({\bf d}) different impurity potential strengths $V_{imp}$ (at fixed
$U=2.5t$). } \label{Fig1}
%{\bf Peter \& BMA: Suggest we modify 1a
%to indicate a critical doping for the SG phase within the frame of
%the diagram. Also, we should include one additional curve added
%to each figure c and d. Since we discuss $T_g > T_c$ in the text,
%it could appropriately be one of these.}
\end{figure}

Hints of magnetic ordering in the SC state had been
detected earlier by $\mu SR$ experiments in zero
field\cite{ChNiedermayer,Panagopoulos02} as a wedge-shaped
extension of the ``spin glass" phase into the SC dome
of the temperature vs. doping phase diagram of cuprates (see Fig.
1a). Lake \textit{et al.}\cite{Lake} reported that an
incommensurate magnetic order similar to the field-induced state
was observed in zero field, too. But although it also vanished at
$T_c$, the ordered magnetic moment in zero field had a $T$
dependence which was qualitatively different from the
field-induced signal. The zero-field signal was attributed to
disorder, but the relation between impurities and magnetic
ordering remained unclear. Because  strong magnetic fluctuations
with similar wavevector are reported at low but nonzero energies
in inelastic neutron scattering experiments on these materials,
e.g. on optimally doped LSCO samples exhibiting  no spin-glass
phase in zero field, it is frequently argued that impurities or
vortices simply ``pin" or ``freeze" this fluctuating
order.\cite{Kivelson}

Describing such a phenomenon theoretically at the microscopic
level is difficult due to the inhomogeneity of the interacting
system, but it is important if one wishes to explore  situations
with strong disorder, where the correlations  may no longer
reflect the intrinsic spin dynamics of the pure system. Such an
approach was proposed by the current authors in a model
calculation for an inhomogeneous $d$-wave SC with
Hubbard-type correlations treated in mean field.\cite{Schmid1} In
this model a single impurity creates, at sufficiently large
Hubbard interaction $U$ and impurity potential strength $V_{imp}$,
a droplet of staggered magnetization with a size corresponding to
the AF correlation length of the hypothetical pure
system.\cite{Tsuchiura,Wanglee,Harter} Such local impurity-induced
magnetism has been studied extensively both theoretically and
experimentally, and was recently reviewed in Ref.
\onlinecite{Alloul}.  As was shown in Ref. \onlinecite{Schmid1},
when these droplets come close enough to interact, there is a
tendency to form incommensurate, phase-coherent N\'{e}el domains
whose size is sufficient to explain the observations by Lake {\it
et al.}\cite{Lake} in zero field. In addition, such a model
explains the empirical observations that both increasing
disorder\cite{Kimura} and underdoping\cite{Panagopoulos02} enhance
the SDW order.

In this paper we investigate the origin of the ``order by disorder"
phenomenon described in Ref. \onlinecite{Schmid1}, as well as the
$T$ evolution of the disordered magnetic state in applied
magnetic field. In the case of the field-induced
SDW, an apparently very natural approach to the problem was
developed by Demler {\it et al.},\cite{Demler} who constructed a
Ginzburg-Landau (GL) theory for competing SDW and SC order in a magnetic field.
%Instead of considering the full solution for the
%inhomogeneous coexistence state in the presence of a field, they
%constructed a theory based on a spatially uniform order parameter
%with effective field-dependent coupling constant
%$g_{eff}=g-A(H/H_{c2})\log (H_{c2}/H)$, where $A$ is a constant of
%order unity.
This order-parameter phenomenology describes correctly the
reduction of condensation energy in the vortex phase of the pure
SC, and leads to a phase diagram qualitatively
consistent with experiments.\cite{SachdevRMP} The GL approach,
however, ignores the energies of the quasiparticles moving in the
inhomogeneous state which also can crucially affect the
competition between SDW and SC order in these materials at low
$T$, as we show here.

In a $d$-wave SC
without AF correlations, a bound state of an isolated
vortex is found at zero energy\cite{MacDonald} due to the sign change of
the order parameter on quasiparticle trajectories through the
vortex core. On the other hand, solutions of the Bogoliubov-de Gennes
(BdG) equations describing coexisting $d$-wave SC and
SDW order\cite{BMAvortex,zhuting,Tingvortex,zhu} show that this
resonance is split by the formation of the SDW; that is, the system
can lower the energy of the nearly bound quasiparticles by moving them
below the Fermi energy. This finding is consistent with the STM
experiments in the Abrikosov state of YBa$_2$Cu$_3$O$_{7-\delta}$
(YBCO) by Maggio-Aprile \textit{et al.} \cite{Maggio-Aprile} and of
BSCCO by Pan \textit{et al.},\cite{Pan} which observed split peaks in
the vortex cores. A similar bound state is associated with non-magnetic
impurities such as Zn in BSCCO and has been also imaged by STM.\cite{PanZn}
It is therefore important to explore the role of quasiparticle bound
states and their coupling to the SDW order to identify the origin of both
types of induced local AF in the SC state. A more detailed
understanding of field-induced order is also highly relevant for the
interpretation of the quantum oscillations observed in recent transport
experiment in high magnetic fields.\cite{Taillefer1,Taillefer2} These
oscillations are possibly due to the formation of Fermi surface pockets as
a consequence of SDW ordering and the concomitant reconstructed bandstructure.
In addition, finding ways to understand the effects of disorder is crucial
in order to reveal the intrinsic AF correlations
present in the underdoped part of the cuprate phase diagram, not
least because the magnetic fluctuations at higher energies may be
responsible for SC itself.

The inhomogeneous mean-field theory presented here for electrons
hopping on a square lattice with a $d$-wave pairing potential and
subject to a Hubbard on-site repulsion $U$, reproduces the essential
aspects of the field-induced spin-glass phase shown schematically in
Fig. \ref{Fig1}{\bf a}. The primary purpose of this analysis is to
model the inhomogeneous SC state and does not intend to
 describe the Mott transition to an insulating state at
half-filling; therefore the doping dependence may not be directly
compared to experiment. On the other hand, if doping is assumed to
be correlated with the  ratio of bandwidth to local Coulomb repulsion
$t/U$ in the model, a phase diagram very much like the one found in
various cuprate materials is obtained, as shown in Figs.
\ref{Fig1}{\bf b-d}.  We consider this a reasonable qualitative approach, since
changes in the Fermi surface of these materials reported by angle
resolved photoemission spectroscopy (ARPES) over the "spin glass"
doping range are small,\cite{Damascelli} and it is therefore
plausible that the primary effect on the electronic structure is
due to the correlation induced band narrowing, as discussed in
Ref. \onlinecite{Schmid1}. The phase diagram we
obtain  in Fig. \ref{Fig1}{\bf b} is thus comparable to the $T-x$ phase diagram shown
in Fig. \ref{Fig1}{\bf a}. The magnetically ordered phase can be enhanced both
by the increase of the correlation strength or stronger disorder
potentials, as shown in Figs. \ref{Fig1}{\bf c,d}.

The problem studied here involves several length scales, in particular
the inter-impurity separation, the inter-vortex separation, the
SC coherence and the magnetic correlations lengths,
which can be difficult to disentangle. We obtain results which
reproduce well the qualitative aspects of the experiment by Lake
{\it et al.},\cite{Lake} but show that some features depend on
nonuniversal aspects of disorder, in particular the process of domain
wall nucleation.
%To the extent that
%disorder- and magnetic field induced SDW order have been discussed
%together at all in the literature, it has been assumed that they
%add independently to the ordered moment.\cite{Lake}  We show here
%that instead magnetic and disorder effects interfere when the
%impurity separation becomes comparable to other length scales.
While disorder- and magnetic-field induced SDW order both add to the
ordered moment, the interference of disorder and magnetic-field effects
is quantitatively significant. The domain wall formation proves
responsible for the distinct $T$ dependences of the field- and
the disorder-induced magnetization. An intriguing aspect of the present
theory is that it also includes a crossover from magnetic
droplets to filamentary stripe-like structures in selected regimes of
hole densities and impurity concentrations. The model therefore offers
a concrete route to describe the physics of the pinning of stripe
correlations in the SC state. This insight may prove
relevant for many experiments in the underdoped cuprates which have
been attributed to stripes.

%Another possibility to create quasiparticle excitations is
%applying an orbital magnetic field. Because cuprate
%superconductors are extreme type II superconductors, already small
%magnetic fields are penetrating the superconductor through
%vortices. It is known for a long time that in $s$-wave
%superconductors of type II a bound state is formed in the vortex
%cores with minimum energy of about $\epsilon_0 \approx
%\Delta^2/(2E_F)$, where $\Delta$ is the energy gap and $E_F$ the
%Fermi energy \cite{Caroli}. This behavior is similar to what is
%theoretically predicted for an uncorrelated $d$-wave
%superconductor,
%%from a simple $d$-wave BCS Hamiltonian,
%except that in scanning tunneling microscopy (STM) experiments a single
%peak has never been found in

%the differential conductance, instead a splitted peak was detected in
%YBa$_2$Cu$_3$O$_{7-\delta}$ by Maggio-Aprile

%\textit{et al.} \cite{Maggio-Aprile} and in
%Bi$_2$Sr$_2$CaCu$_2$O$_{8+\delta}$ by Pan \textit{et al.} \cite{Pan}.

%%which we assume to be a fingerprint of antiferromagnetic order inside
%the vortices.
%On the other hand it is well known that the electrons in cuprate
%superconductors are strongly correlated, leading to
%antiferromagnetism, observed by neutrons in vortex cores and their
%surroundings \cite{Lake}.
%%That the vortices in high-$T_c$
%%materials are magnetized was found in neutron scattering experiments.
%Moreover antiferromagnetism was detected in underdoped LSCO even
%in zero field.\cite{Lake}
%%antiferromagnetic order was measured by neutrons even in zero field.

The basis for our model analysis is the BCS pairing Hamiltonian for a
$d$-wave SC with orbital coupling to an applied magnetic
field, to which we add site-centered chemical disorder and a local
Hubbard repulsion; the latter is treated in an unrestricted Hartree-Fock
approximation:
\begin{eqnarray}
H = &-& \sum_{ij\sigma} t_{ij} \: e^{{\rm i} \varphi_{ij}}\: c^{\dagger}_{i\sigma}
c^{}_{j\sigma} - \mu \sum_{i\sigma} c^{\dagger}_{i\sigma} c^{}_{i\sigma}
+\sum_{\langle ij \rangle} \left(\Delta_{ij} c^{\dagger}_{i\uparrow}
c^{\dagger}_{j\downarrow} + h.c.\right)
\nonumber \\
&+& \frac{U}{2} \sum_i\left(\langle n_i \rangle n_i - \langle \sigma_i^z
\rangle \sigma_i^z \right) + \sum_{i\sigma} V_i^{imp} c^{\dagger}_{i\sigma}
c^{}_{i\sigma}.
\end{eqnarray}
Here, $c^\dagger_{i\sigma}$ creates an electron on a square lattice
site $i$ with spin $\sigma=\uparrow,\downarrow$.  The hopping matrix
elements between nearest and next-nearest neighbor sites $i$ and $j$
are denoted by $t_{ij} = t$ and $t_{ij} = t'$, respectively. An
electron moving in the magnetic field from site $j$ to $i$ acquires
additionally the Peierls phase $\varphi_{ij}=(\pi/\Phi_0)
\int^{{\bf r}_i}_{{\bf r}_j} {\bf A}({\bf r})\cdot{\rm d}{\bf r}$, where
$\Phi_0 = hc/(2e)$ denotes the superconducting flux quantum and ${\bf
A(r)} = B (0, x)$ is the vector potential of the magnetic field in the
Landau gauge. The chemical potential $\mu$ is adjusted to fix the
average charge density $n = \frac{1}{N} \sum_i \langle n_i \rangle = 1 - x$,
where $x$ is the hole concentration and $N$ is the total number of
lattice sites; in the following we will focus on $x=0.1$. The magnitude
of the $d$-wave pairing amplitude $\Delta_{ij}$ is determined by the
strength of an attractive nearest-neighbor interaction $V_d$. The
non-magnetic impurity potential $V_i^{imp}$ is described by a set of
pointlike scatterers at random positions, and all fields, i.e. the pairing amplitude 
$\Delta_{ij}$, the local charge density $\langle n_i \rangle$, and the local 
magnetization $\langle \sigma_i^z
\rangle$ are calculated self-consistently from the solutions of the
associated BdG equations. (For further details of the numerical method
we refer to the Supplementary Information.)

%We determine the local density of states (LDOS), which is proportional
%to the spatially resolved differential conductance measured in STM
%experiments. For a direct comparison with neutron scattering data we
%evaluate the Fourier transformed magnetization $M({\bf q})$, because in
%the approximation of a factorized spin-spin correlation function
%$|M({\bf q})|^2$ becomes equal to the magnetic structure factor.

%{\bf This paragraph either needs to be expanded to include
%definitions of M(r), M(q), and LDOS, or it should be eliminated
%altogether; it's too short and interrupts the flow of the
%ms.--Peter.}

\begin{figure}[t!]
\vspace{.5cm}
\hspace{-.5cm}
\begin{overpic}[scale=.28,unit=1mm]{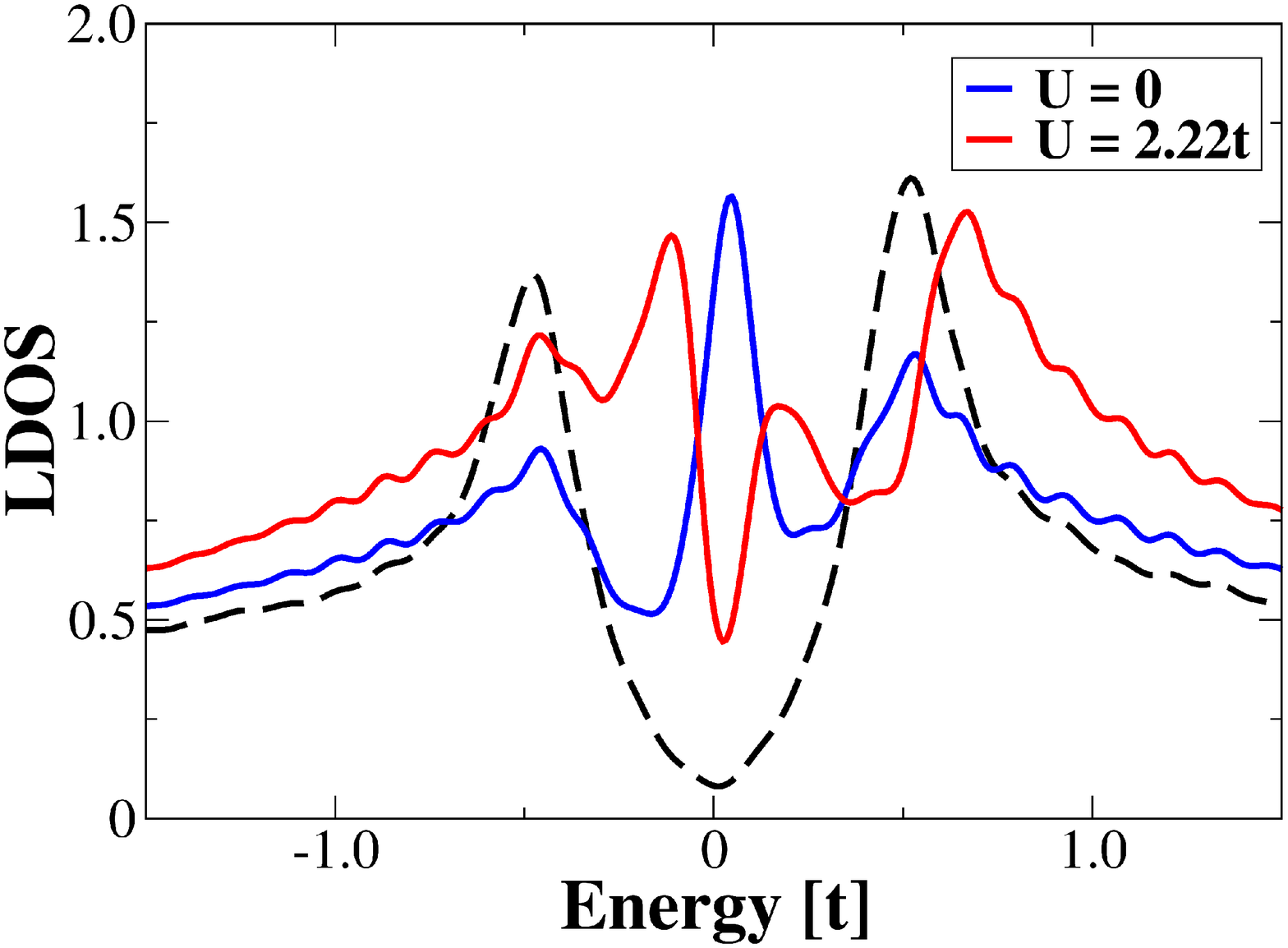}
    \put(-3,48){{\bf a}}
 \end{overpic}
\hspace{1cm}
\begin{overpic}[scale=.28,unit=1mm]{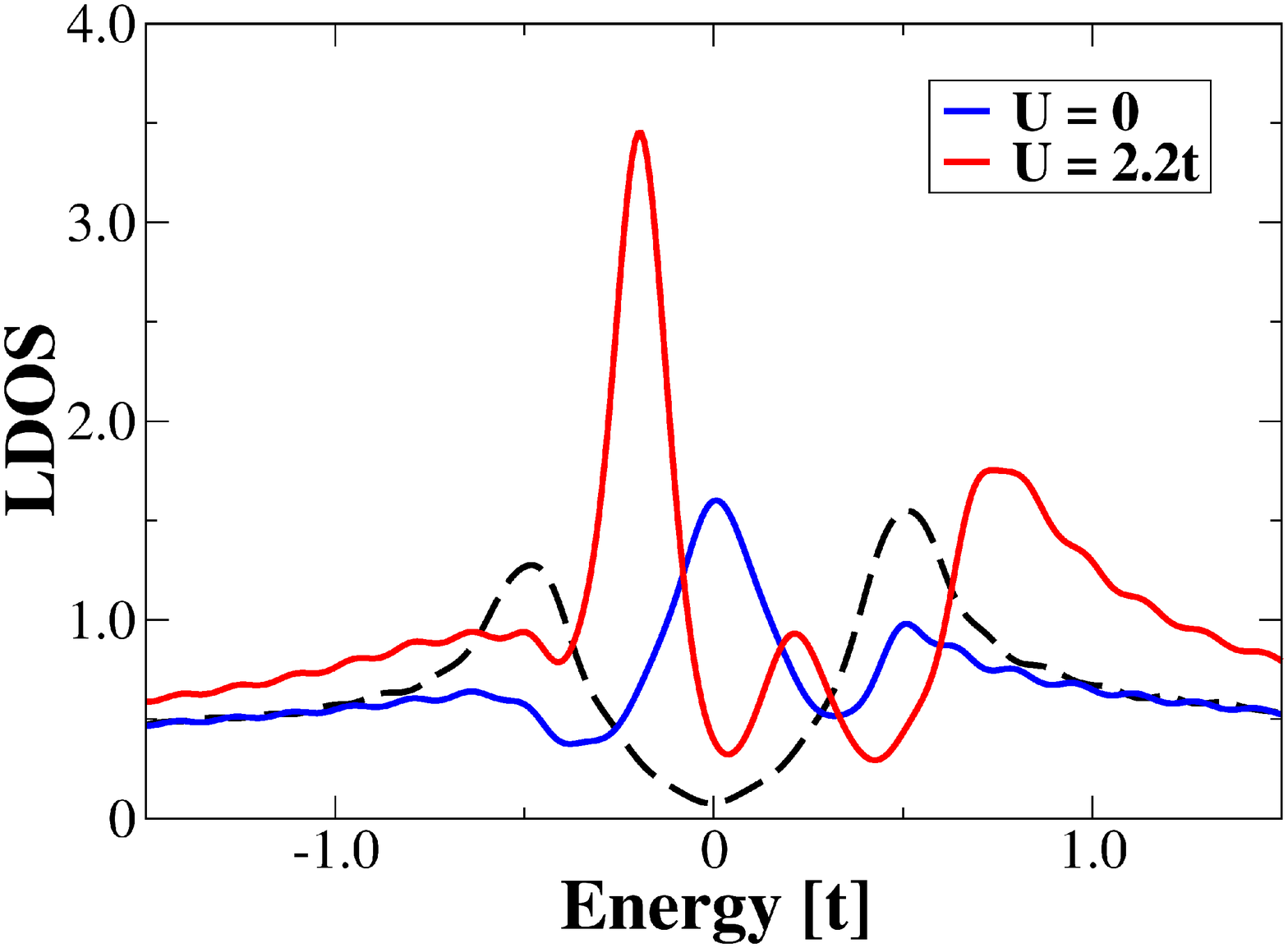}
    \put(-3,48){{\bf c}}
 \end{overpic}
%\includegraphics[width=7.5cm]{fig1d.eps}

%\hspace{.3cm}
 \begin{overpic}[scale=.98,unit=1mm]{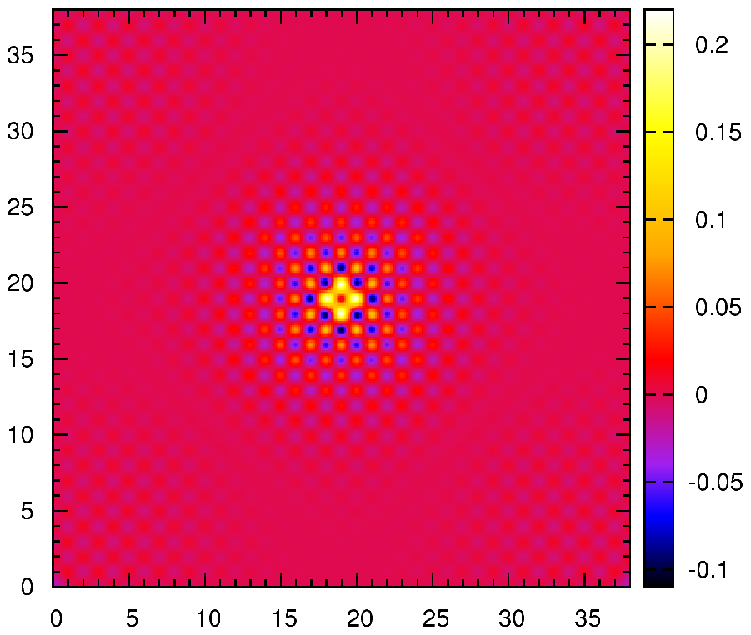}
    \put(-3,61){{\bf b}}
 \end{overpic}
\hspace{0.5cm}
 \begin{overpic}[scale=.98,unit=1mm]{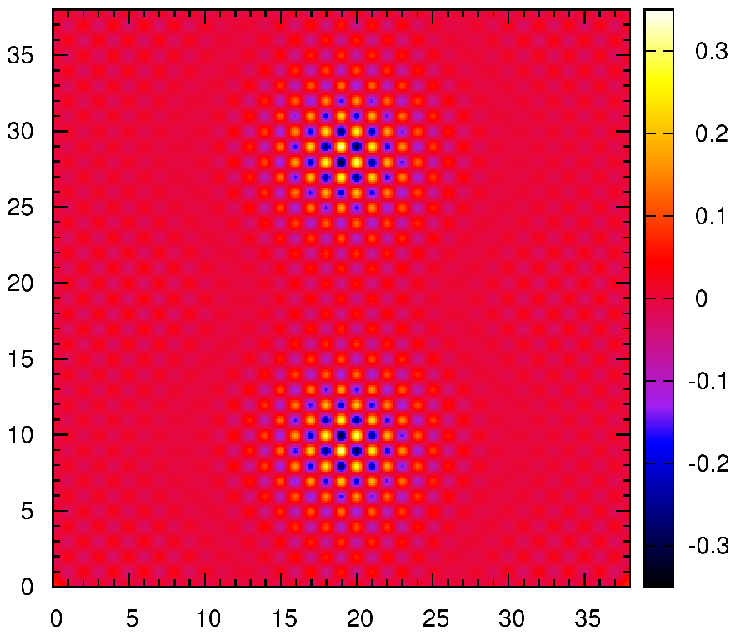}
    \put(-3,61){{\bf d}}
 \end{overpic}
\caption{\setlength{\baselineskip}{5mm}
 {\bf Impurity- and field-induced magnetization.} {\bf a}, {\bf c}. Local
density of states (LDOS) at a nearest-neighbor site of a single impurity
({\bf a}) and in the center of one vortex ({\bf c}). In both cases, above
a critical $U_c$, local SDW order is induced, concomitantly the zero-energy
peak in the absence of the Hubbard repulsion (blue curve) splits
spin-dependently (red curve). The dashed curves show the clean LDOS far
from the perturbation. {\bf b}, {\bf d}. Real-space patterns of the
magnetization $\langle \sigma_i^z \rangle$ in units of $\mu_B$ on a
$38\times38$ lattice for $U=2.2t > U_c$
at low temperature $T=0.025t$. {\bf b} shows the magnetization nucleated
by a strong impurity ($V_{imp} = 60t$) located at the center. In {\bf d}
two superconducting flux quanta $\Phi = 2\Phi_0$ thread an impurity-free
$d$-wave SC.}
%{\bf Can we enhance
%contrast of current 2a and 2c? Perhaps a different color scheme is better.}
\label{Fig2}
\end{figure}

We start with a single non-magnetic impurity in a $d$-wave SC at $U=0$. 
The fingerprint of the
induced virtual bound state is a single near zero-energy peak in the
local density of states (LDOS) (see Fig. \ref{Fig2}{\bf a}). In this
situation there is no magnetization induced by the impurity.
Increasing the on-site Coulomb repulsion beyond a critical value
$U_c$, a staggered magnetization emerges in the neighborhood of
the impurity (Fig. \ref{Fig2}{\bf b}). The two-sublattice nature of
the magnetic pattern, in conjunction with the spatial extent of the
impurity-induced resonance leads to a spin-dependent splitting of the
near-zero-energy peak in the LDOS; one resonant state with a
selected spin direction is thereby shifted below and the other
above the Fermi energy, as is most clearly seen at the
nearest-neighbor sites of the impurity where the resonant state
has the largest weight (see Fig. \ref{Fig2}{\bf a}). The spin-dependent
splitting reduces the bound-state energy of a spin-up or -down
state, and therefore stabilizes a local  droplet of staggered
magnetization which carries a total spin 1/2-moment surrounding the
impurity.
%divides the lattice into
%two sublattices with opposite magnetization. This doubling of the
%magnetic unit cell halves the Brillouin zone and therefore leads to a decrease in energy for the less than
%half filled case. Hence the energy of an electron on a given site depends on the spin.
%This behavior can be seen in the LDOS at the next-neighbor site to the impurity.
%With increasing $U$ the resonance peak starts to spli
%spin-dependently and
%spectral weight of a spin-up or a spin-down state, respectively,
%is shifted below the Fermi energy, which reduces the bound state energy, and leads to a local magnetization
%in a halo surrounding the impurity.
The splitting of the resonance peak can therefore be viewed as the
origin of the impurity-induced magnetization. It is
important to note that the splitting of the bound state is {\it
not} due to the suppression of the $d$-wave order parameter near
the impurity. As we have verified,  the order parameter can be
artificially held constant in the solution of the BdG equations, and a nearly
identical result is obtained.

%\begin{figure}[t!]
%\hspace{1em}
%\includegraphics[width=7.5cm]{fig3a.eps}
%\hspace{0.2em}
%\includegraphics[width=7.5cm]{fig3b.eps}
%\includegraphics[width=7.5cm]{fig3c.eps}
%\caption{\setlength{\baselineskip}{5mm}
%{\bf Magnetic structure factor of a disordered system} (10\% impurities, $V_{imp} = 1.3t$, $T=0.025t$). ${\bf q}$-space
%plot {\bf a} in zero-field and {\bf b} at finite-field. {\bf c}, Temperature-dependence of the summed incommensurate
%peaks of the magnetic structure factor in zero-field (blue curve) and at finite field with the zero-field data
%subtracted (red curve).}
%\label{Fig3}
%\end{figure}

In the presence of a thermodynamically finite density of impurities, we
recover at $T=0$ the results of  Ref. \onlinecite{Schmid1}, i.e. the
creation of a defected but magnetically ordered state as defined by
strong peaks in the Fourier transform of the local magnetization $M(\q)$
at incommensurate wavevectors ${\bf q}$ near $(\pi,\pi)$. The effect of
temperature is now naturally included in the theory via $T$-dependent occupation
probabilities of Bogoliubov quasiparticle states, as discussed in detail
in the Supplementary material. In Fig. \ref{Fig1}{\bf b} we show the
extent of the quasi-ordered phase, labelled ``spin glass", which expands
as correlations increase, and disappears at a critical value of $t/U$
which depends on the strengths of the pairing interaction and the
impurity potential. The intensity of the incommensurate magnetic Bragg
peaks is shown in Figs. \ref{Fig1}{\bf c} and \ref{Fig1}{\bf d} as
functions of $T$ for fixed $t/U$ and fixed impurity potential
strength, respectively. From Fig. \ref{Fig1}{\bf c} it becomes evident
that the magnetic ordering, or ``glass transition" temperature $T_g$ can
be smaller, equal to, or larger than $T_c$ depending on $U$.
%but that the amplitude of magnetic order is suppressed in
%the normal state relative to the superconducting state if
%$T_g>T_c$ (Can we include a curve like this? Peter). This is
%because within the present theory there is no gap in the spectrum
%above $T_c$ to promote bound state formation.
In Fig. \ref{Fig1}{\bf d} we show how increasing the impurity potential
$V_{imp}$ can increase both the amplitude of the disorder-induced SDW
and $T_g$ itself. These results are consistent with the empirical
observation that the size of the spin-glass phase is not universal, and in
particular the critical doping beyond which magnetic order is no longer
observed varies considerably between intrinsically disordered cuprates like LSCO and BSCO,
 and
clean materials like YBCO.

\begin{figure}[t!]
\includegraphics[width=7.5cm]{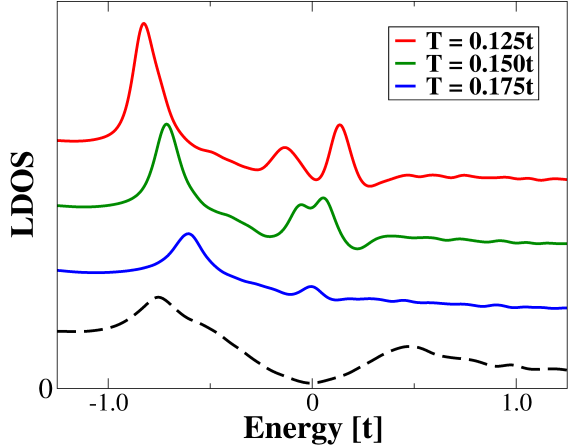}
\caption{\setlength{\baselineskip}{5mm}
{\bf Temperature dependent peak splitting in a magnetic field}.
LDOS at the vortex center for three different temperatures below $T_c$ in
the absence of impurities.  Below the
critical temperature $T_g = 0.175t$, the vortices magnetize and simultaneously the  Andreev bound-state peak splits.
The black dashed curve shows the LDOS far away from the vortex.}
%{\bf BMA: Is the figure important enough to include as a standalone figure?
%The reader have to just believe us that magnetization in not present
%in the blue curve of the present format of this figure. We could consider
%"sexying up" this figure by including smaller panels on the side of the
%magnetization in real-space. Alternative an idea would be to make the
%real-space patterns in Fig. 2b,d insets in the LDOS figures 2a,c, and
%incorporate this Fig. 3 as a third subfigure in Fig. 2.}
\label{Fig3}
\end{figure}

The magnetization induced by an orbital magnetic field can be traced to the
same microscopic origin as  the impurity-induced
magnetization.\cite{zhuting}
%In Fig.~1c we see two magnetic flux quanta %$\Phi_0 = \frac{hc}{2e}$
%are penetrating the superconductor through vortices with their centers located at sites $(19a, 9a)$ and $(29a, 9a)$.
%In Fig.~2c we observe two magnetized vortices
Above a critical $U_c$ a staggered spin pattern is nucleated in the vortex
cores with a spatial extent reaching  beyond the size of a vortex core (see
Fig. \ref{Fig2}{\bf d}), as  observed in experiment.\cite{Lake} For the
parameter set chosen, the core radius estimated from the area where the
order parameter is suppressed is about one lattice spacing.
%the antiferromagnetic order extends significantly beyond that. %While the order parameter is suppressed
%substantially only on one lattice site, the antiferromagnetic order exists over many lattice constants.
The LDOS in the vortex center reveals that the origin of the field-induced
magnetization is tied to the spin-dependent splitting of the Andreev bound
state in the vortex core (Fig. \ref{Fig2}{\bf c}). The conjecture that the
field-induced magnetization  indeed appears simultaneously with the peak
splitting in the LDOS is explicitly verified in Fig. \ref{Fig3}. For the
unmagnetized vortex at $T=0.175t$ a single Andreev bound-state peak exists
at zero energy. With decreasing $T$ the vortices nucleate a
staggered spin pattern precisely at the $T$ where the zero-energy
peak in the LDOS splits. With further cooling the peak splitting grows, 
more spectral weight is shifted  below the Fermi energy, and the
magnetization is  enhanced.

\begin{figure}[t!]
\hspace{.3cm}
 \begin{overpic}[scale=.91,unit=1mm]{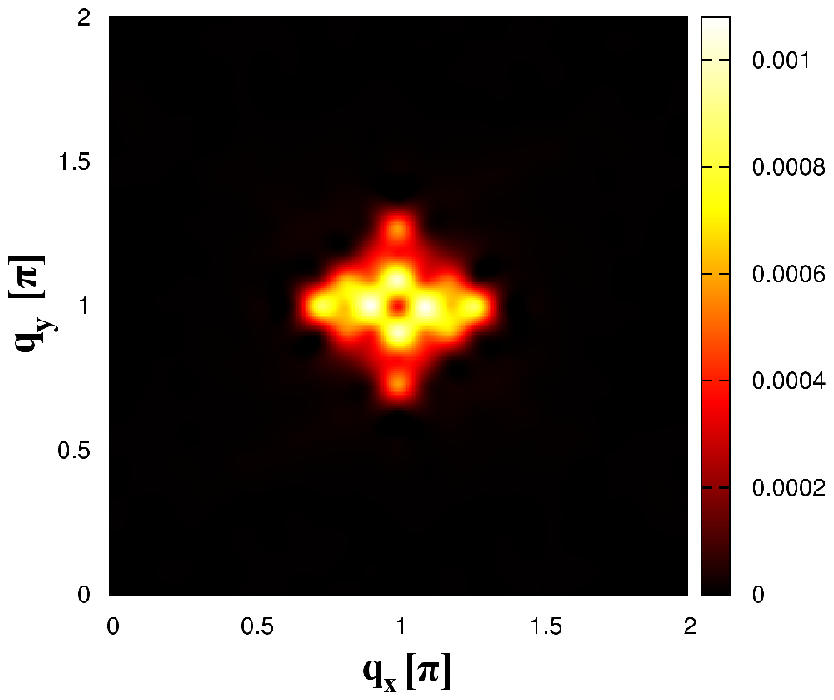}
    \put(-2,59){{\bf a}}
 \end{overpic}
\hspace{0.5cm}
 \begin{overpic}[scale=.91,unit=1mm]{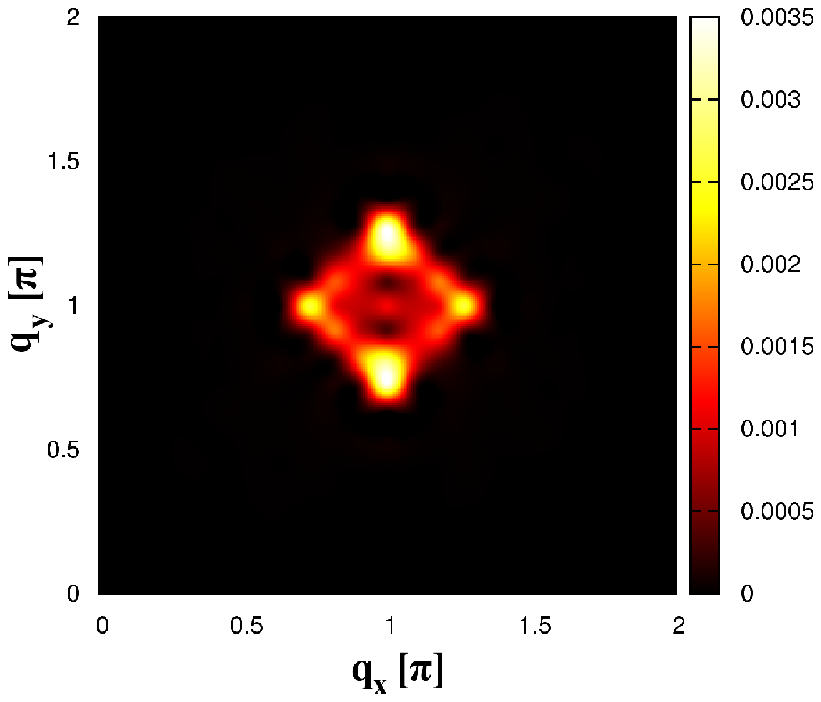}
    \put(-2,59){{\bf b}}
 \end{overpic}

\vspace{0.5cm}
 \begin{overpic}[scale=.29,unit=1mm]{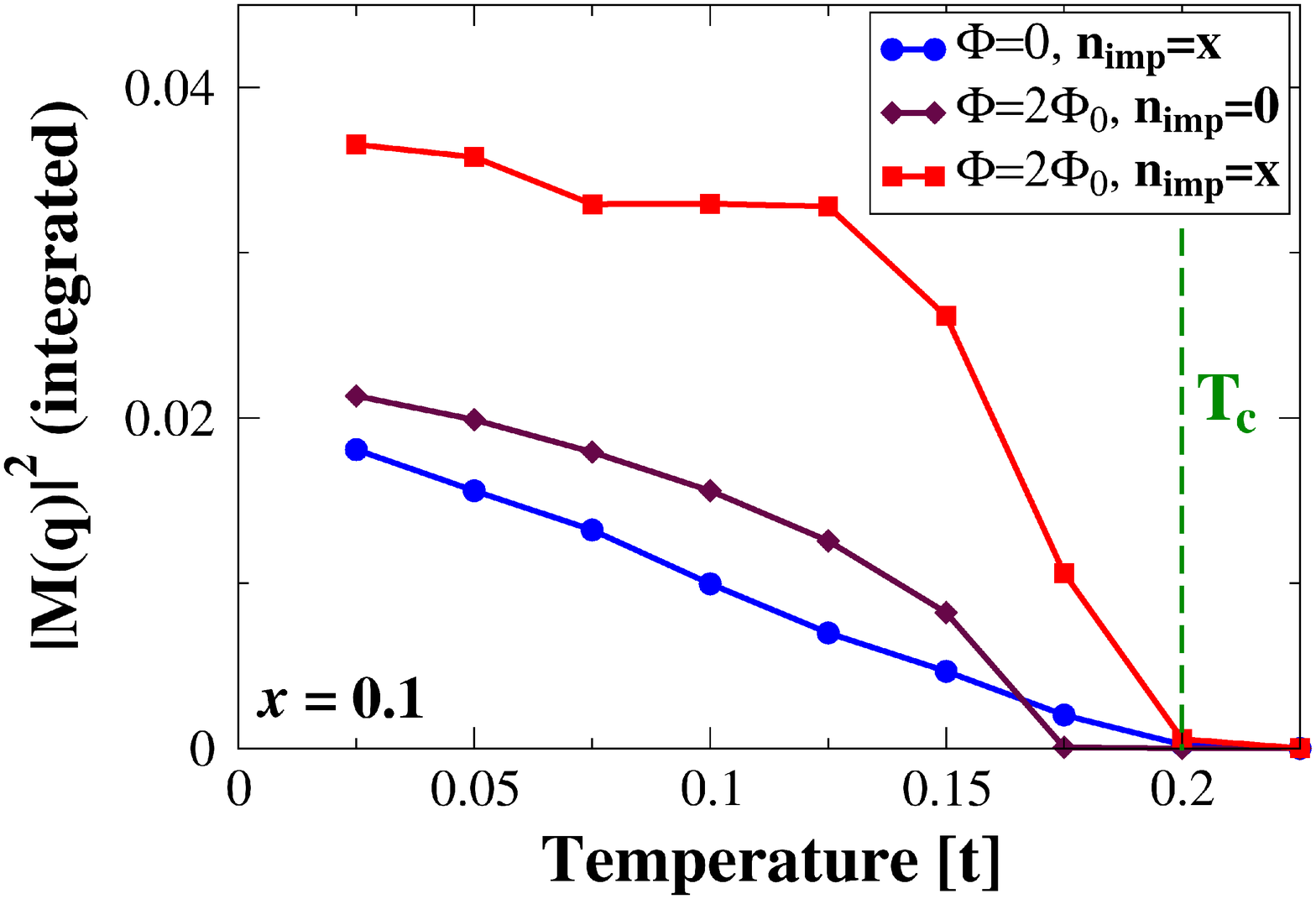}
    \put(-3,47){{\bf c}}
 \end{overpic}
\caption{\setlength{\baselineskip}{5mm} {\bf Averaged magnetic
structure factor for a many-impurity system.} {\bf a,b}. Intensity
plot of the magnetic structure factor around $(\pi,\pi)$ at
$T=0.025t$ in zero magnetic field ({\bf a}) and at finite field
({\bf b}). The structure factor data were averaged over ten different
impurity configurations. For the used system size of $22 \times 22$
lattice sites a magnetic flux of $2\Phi_0$ corresponds to a strong
magnetic field with $H = 59T$. The impurity concentration $n_{imp}=x$
is fixed to $10\%$ ($V_{imp} = 1.3t$) and $U = 2.9t > U_c$. {\bf c}.
$T$-dependence of the peak intensity integrated around $(\pi,\pi)$
in zero-field and at finite field with the finite density of impurities
$n_{imp}=x$ (blue and red curve, respecctively); for the data with
$\Phi=2\Phi_0$ and $n_{imp}=x$ the zero-field data were subtracted.
For comparison also the structure factor in a clean system is included
for the same magnetic-field strength (purple curve). $|M({\bf q})|^2$
(integrated) translates directly to the ordered spin moment squared in
units of $\mu_B$ per Cu$^{2+}$.}
%{\bf Peter:  Since we argue that the field- and
%disorder-induced signals do not add independently, does it make
%sense to show only the curve with zero field subtracted a la Lake?
%Perhaps we should have absoulte scales, no subtraction with the
%subtracted curve in an insert?  2nd, I would like to see a more
%pixelated, less smoothed version of the M(q) patterns. Also, Arno,
%you said we could potentially get a curve for smaller field to
%show here. Did Markus start this?}
\label{Fig4}
\end{figure}

The natural next step is to consider a finite density of
non-magnetic impurities in the presence of an external magnetic
field and to compare it to zero-field results.  Specifically for
the modelling of LSCO, we assume in the following that the Sr ions
are the primary source of disorder, such that $n_{imp} = x$, where
$n_{imp}$ denotes the impurity concentration. These systems are in
the strongly disordered regime where the AF correlation length (droplet size) is
comparable to the average distance between the dopants, such that
the disorder is far from the one-impurity limit. Since the Sr
dopants are removed from (but close to) the CuO$_2$ planes, we
model them as weak scatterers with $V_{imp}
= 1.3t$. Figs. \ref{Fig4}{\bf a,b} show the magnetic structure factor
$S({\bf q})$ at a fixed temperature far below $T_c$ in zero and
finite magnetic field averaged over ten different impurity
configurations. As  in Figs. \ref{Fig1}{\bf c,d}, $S({\bf q})$ is
approximated by $|M({\bf q})|^2$ with $M({\bf
q})=\frac{1}{N}\sum_i e^{i {\bf q \cdot r}} \langle \sigma_i^z
\rangle$ (see the Supplementary information for details).
The magnetic signal in the structure factor appears at the
incommensurate wavevectors ${\bf q}= (\pi,\pi\pm\delta)$ and ${\bf
q}=(\pi\pm\delta,\pi)$ (see Fig. \ref{Fig4}{\bf a}). The magnitude of the
incommensurability $\delta$ however varies for distinct impurity
configurations randomly selected for $22 \times 22$ lattice
systems. For the weak magnetic signal in zero field the averaging
over different impurity configurations is therefore imperative but
computationally demanding. Applying an external magnetic field
strongly enhances the magnetization and reinforces incommensurate
peaks at unambiguously selected wavevectors which are robust against
variations in the impurity configurations (See Fig. \ref{Fig4}{\bf b}).

Remarkably, the temperature dependence of the structure factor (see Fig.
\ref{Fig4}{\bf c}) closely resembles the
neutron scattering data on LSCO by Lake {\it et al.}\cite{Lake}
For the results shown in Fig. \ref{Fig4} we have chosen a parameter set
where the staggered magnetization in zero field has its onset at a
temperature $T_g$ indistinguishable from $T_c$. This
reflects a situation where upon cooling through $T_c$ the
localized bound states inside the $d$-wave energy gap emerge and
immediately split in the self-stabilizing staggered magnetic
pattern. Towards lower $T$ the magnetic structure factor
rises in a markedly different way in zero and in finite field.
While the field-induced part of the magnetic signal has a negative
curvature, the zero-field magnetic structure factor increases
approximately linearly upon cooling. The two mechanisms of impurity-
and field-induced SDW do  not simply cooperate
additively; the field-induced part of the magnetization is twice as
large in the presense of impurities as compared to the field-induced
SDW in the clean system (see Fig. \ref{Fig4}{\bf c}).
The zero-field increase of the magnetization in the inhomogeneous
SC state originates from the merging of AF
patches nucleated by the individual impurities. Without impurities the
increasing field-induced magnetization with cooling results from the
growth of the magnetized regions around the well-separated vortex cores.
Thus, both our zero- and finite-field results for the finite density
of non-magnetic impurities $n_{imp}=x$ closely follow form of the $T$ dependence
of the the neutron
scattering data for underdoped LSCO.\cite{Lake} Still, due to
computational restrictions we are not yet able to access the low
magnetic-field strengths to allow for a direct comparison with experiment.

\begin{figure}[t!]
\hspace{1em}
\begin{overpic}[scale=1.0,unit=1mm]{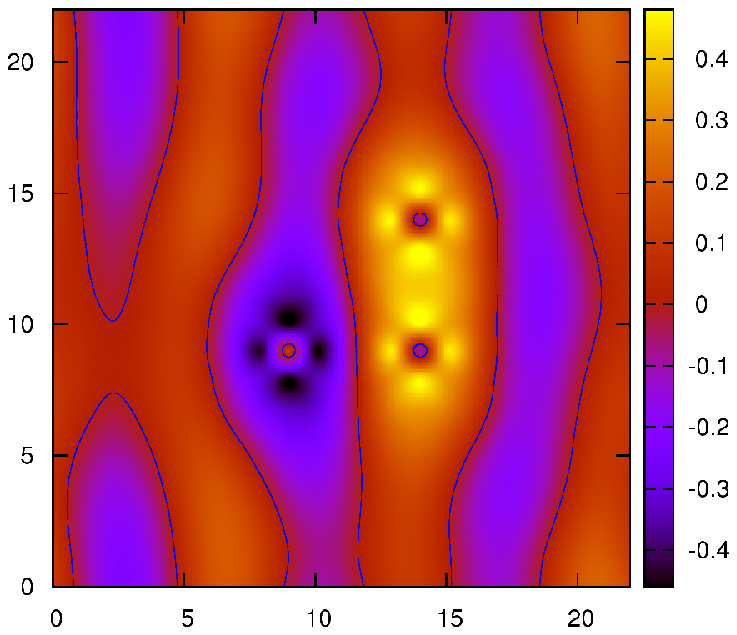}
    \put(-3,61){{\bf a}}
 \end{overpic}
\hspace{0.5cm}
\begin{overpic}[scale=1.0,unit=1mm]{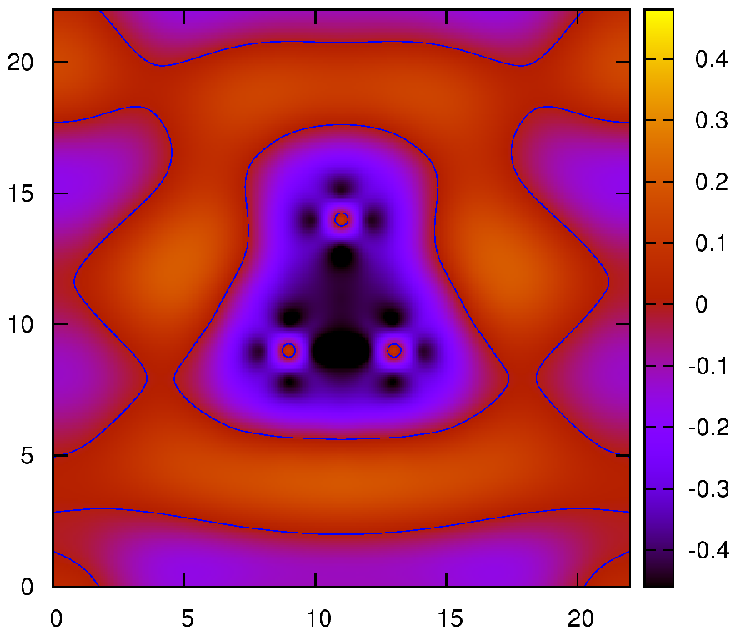}
    \put(-3,61){{\bf b}}
 \end{overpic}

\vspace{0.5cm}
 \begin{overpic}[scale=.29,unit=1mm]{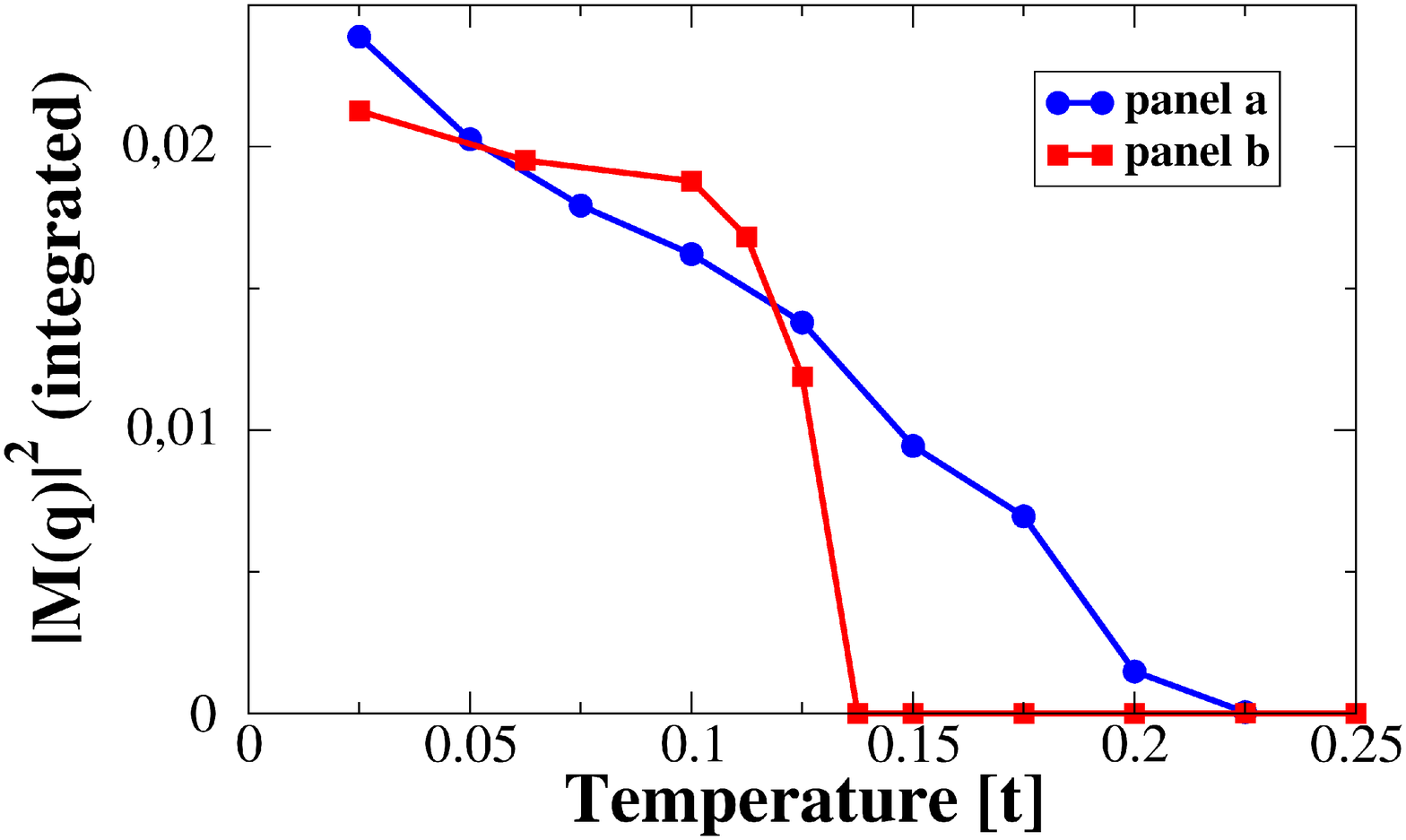}
    \put(-3,47){{\bf c}}
 \end{overpic}
\caption{\setlength{\baselineskip}{5mm} {\bf Anti-phase domain walls.
a, b.} Real-space image of the staggered magnetization at $T=0.025t$
induced by three non-magnetic impurities. In impurity
configuration {\bf a} anti-phase domain walls appear vertically.
In configuration {\bf b} the staggered magnetization induced by
the three impurities adjusts to a uniform AF domain
around them. {\bf c}. Temperature dependence of the integrated
magnetic structure factor for the impurity configurations {\bf a}
(blue curve) and {\bf b} (red curve), respectively.}
%{\bf What do these figures look like without smoothing?}
\label{Fig5}
\end{figure}

An important remaining question is why the $T$
dependences of the magnetization are different in zero and in
finite field. A hint is provided by the observation that the
$T$ dependence of the magnetic structure factor for the
impurity-free field-induced magnetization and also for just two single
impurities in zero field has a negative curvature. In both cases the
induced staggered magnetization patterns around each impurity or each
vortex, respectively, adjust their individual two-sublattice spin
structures in phase and thereby avoid any domain walls.\cite{Shender}
For three nearby impurities, however, it proves already difficult to
find a specific configuration where anti-phase domain walls are
absent. In Figs. \ref{Fig5}{\bf a,b} we compare the staggered
magnetization of two 3-impurity configurations with distinctly
different domain-wall patterns. Remarkably, placing the three
impurities on the same sublattice to form a right-angled
equilateral triangle as in Fig. \ref{Fig5}{\bf a} generates a sequence of
vertical anti-phase domain walls. If instead the impurities are
configured in an acute equilateral triangle as in Fig. \ref{Fig5}{\bf b} a
simply connected AF island forms around them. As
Fig. \ref{Fig5}{\bf c} shows, with decreasing $T$ the magnetic
signal evolves differently for each impurity configuration.
Intriguingly, $|M({\bf q})|^2$ rises almost linearly for the
configuration with vertical anti-phase domain walls while it has a
negative curvature for the single domain island. These examples, and others we have investigated,
suggest that the linear low-$T$ rise of the
magnetic signal for a finite density of impurities in zero field
originates from the anti-phase domain walls which are always present
in the randomly generated impurity configurations. For the field-induced
magnetization  it is the larger distance between the
magnetized vortices which prevents the occurrence of domain walls
and therefore alters the $T$ dependence of $|M({\bf q})|^2$.

All the results presented above focused on static disorder- and
field-induced SDW, but inelastic neutron scattering
experiments have shown that in the SC state the spin
excitations at finite energy have almost the same distribution of
spectral weight in $\bf q$ as the frozen magnetic order.\cite{tranquadareview}
For very low doping in the normal spin glass phase above $T_c$, the neutron
intensity pattern is rotated by 45$^\circ$ and the connection to the
spin correlations discussed here is less obvious. In fact, the utility of our
model of choice is questionable for the description of the normal state
where Fermi liquid concepts may not even be applicable. Nevertheless in
the SC state we have provided a concrete foundation for the
freezing of fluctuating spin correlations by disorder and magnetic field
on the same footing; in particular the role of the quasiparticle bound states
in the formation of the magnetic order has been highlighted.

The new picture that emerges is complementary to the global competition between SC and AF phases
in the sense that SC and disorder may significantly enhance SDW order in the underdoped regime. The $d$-wave
pairing of the SC condensate is crucial for this generation of local magnetism, as we have shown. Support for this cooperative effect
between SDW and SC comes not only from the onset of the elastic magnetic neutron signal at $T_c$ but
also from  Zn-substituted optimally doped LSCO. There it is found by $\mu$SR that 2\% Zn 
induces a magnetic signal, but $3\%$ Zn is found to eliminate it, but also destroys superconductivity\cite{Watanabe}; within the context of the current theory, this effect is understood 
 not as a consequence of spin dilution\cite{Watanabe}, but rather due to the destruction of the SC phase and thereby its ability to generate (or enhance) magnetic order.

\begin{figure}[t!]
\hspace{1em} \begin{overpic}[scale=1.0,unit=1mm]{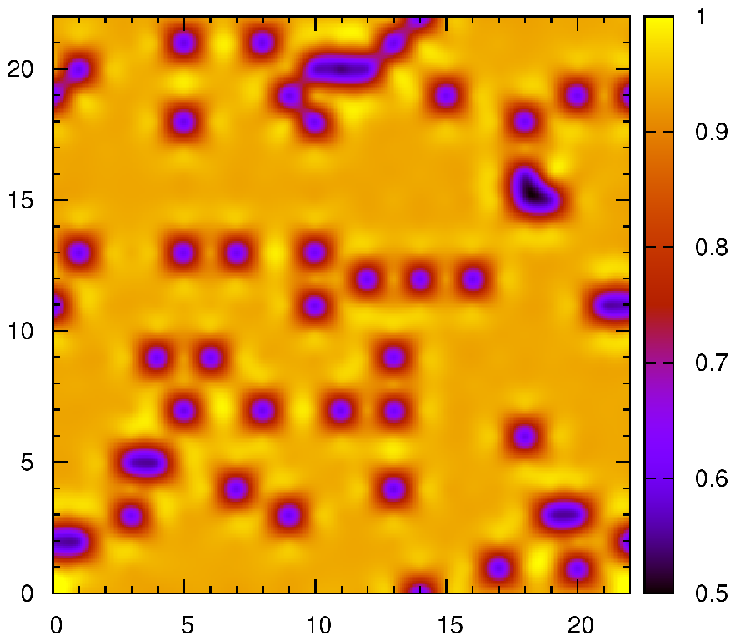}
    \put(-3,61){{\bf a}}
 \end{overpic}
\hspace{0.5cm}
\begin{overpic}[scale=1.0,unit=1mm]{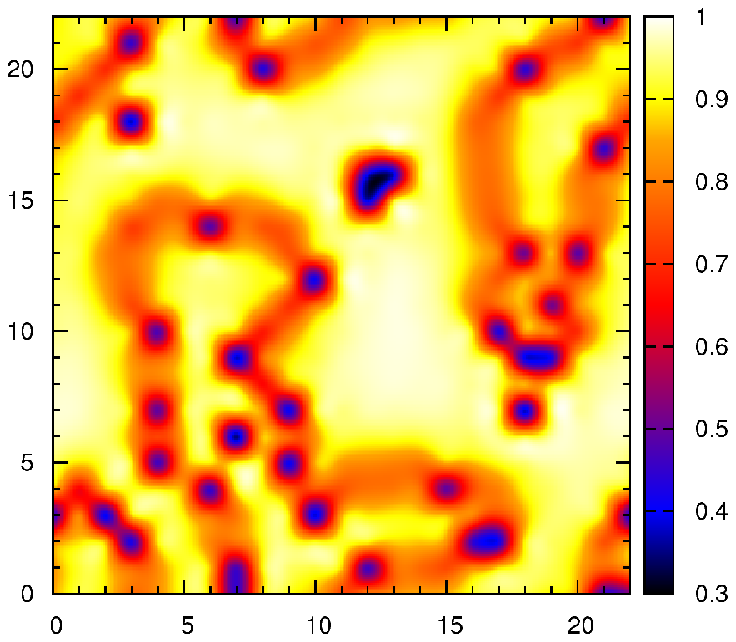}
    \put(-3,61){{\bf b}}
 \end{overpic}
%\vspace{-1cm}
%\includegraphics[width=12cm]{fig6a.eps}
%\vspace{-1cm}
\caption{\setlength{\baselineskip}{5mm} {\bf Charge-density profiles.}
{\bf a}. The same parameter set in zero field as in Fig. 4: $U=2.9t,
V_{imp}=1.3t$, doping $x=10\% =n^{imp}$, and pairing interaction strength
$V_d=1.34t$. {\bf b}. $U=4.0t, V_{imp}=1.3t$, doping $x = 15\%$, $n^{imp}=
7.5\%$, and $V_d=2.0t$. In ({\bf a}) and ({\bf b}) the
temperature is $T=0.025t$.}
%{\bf Need to find sequence of 3 $U$'s and all other parameters fixed which
%shows the patches-to-stripes transition very clearly.}
\label{Fig6}
\end{figure}

Finally we show that a qualitatively different kind of inhomogeneous
textures may also be stabilized within the present weak-coupling approach.
Figure \ref{Fig6}{\bf a} shows the typical charge-density profile in zero
field for a parameter set used above to explore the onset of static
AF. As expected, at and near the impurity sites the electron
density is reduced, and in these areas the local SDW patches nucleate. With
increasing repulsion $U$ and for larger hole densities exceeding the impurity
cocncentration, the inhomogeneous spin and charge patterns change
qualitatively, and the impurity-centered patches with reduced electron
density evolve into hole-rich filamentary structures (see Fig.
\ref{Fig6}{\bf b}). In this still SC state the filaments
constitute snake-like paths through an SDW background with an average
density of almost one electron per site. These textures provide a link to
the study of disordered (quenched) stripes similar to those discussed
recently within various GL models.\cite{robertson,delmaestro,vojtareview}
Therefore, depending upon the correlation strength and the details of the
disorder, the magnetic ordering temperature $T_g$ can vary significantly,
and the ordering itself can be droplet-like or filamentary-like. This may
explain much of the variability of neutron and $\mu SR$ experiments on
different cuprates. Many interesting open questions remain to be addressed
in future work, including the possibility of nematic instabilities in the
presence of weak symmetry breaking fields and the transfer of
spin-fluctuation spectral weight to finite energies in samples where
magnetism is not frozen.

\appendix
\section{Numercial Method}
In order to investigate disorder- and field-induced magnetic order in $d$-wave superconductors we self-consistently
solve the Bogoliubov - de Gennes (BdG) equations on  a square lattice for the Hamiltonian
\begin{eqnarray}
  H = &-& \sum_{ij\sigma} t_{ij} \: e^{{\rm i} \varphi_{ij}}\: c^{\dagger}_{i\sigma} c^{}_{j\sigma} -
       \mu \sum_{i\sigma} c^{\dagger}_{i\sigma} c^{}_{i\sigma}
      +\sum_{\langle ij \rangle} \left(\Delta_{ij} c^{\dagger}_{i\uparrow} c^{\dagger}_{j\downarrow} + h.c.\right)
       \nonumber\\
      &+& \frac{U}{2} \sum_i\left(\langle n_i \rangle n_i - \langle \sigma_i^z \rangle \sigma_i^z \right) +
      \sum_{i\sigma} V_i^{imp} c^{\dagger}_{i\sigma} c^{}_{i\sigma},
\label{eq:h}
\end{eqnarray}
where %$t_{ij}$ describes
the hopping amplitude between nearest neighbor and next-nearest neighbor sites $i$ and $j$ is described by
$t_{ij}=t$ and $t_{ij}=t' = -0.4t$, respectively. An orbtial magnetic field is represented by the Peierls phase factor
$\varphi_{ij}= (\pi/\Phi_0) \int^{{\bf r}_i}_{{\bf r}_j} {\bf A}({\bf r})\cdot{\rm d}{\bf r}$,
while  $\Phi_0 = hc/(2e)$ is the superconducting flux quantum and ${\bf A(r)} = B (0, x)$ is the vector potential of
the magnetic field in the Landau gauge. The $d$-wave pairing potential is defined on two nearest neighbor sites 
$i$ and $j$ by
\begin{eqnarray}
\Delta_{ij} &=& -V_d \langle c_{j\downarrow} c_{i\uparrow}\rangle = \Delta_{ji}
\label{eq:op1},
%n_{i\sigma} &=& c_{i\sigma}^{\dagger} c_{i\sigma}^{}  \label{eq:cd}\\
%n_i &=& n_{i\uparrow} + n_{i\downarrow} \\
%\sigma_i^z &=& n_{i\uparrow} - n_{i\downarrow}. \label{eq:sd}
\end{eqnarray}
where $V_d$ is the attractive pairing interaction strength, which we set to $V_d = 1.34t$ throughout the paper.
%Singlet pairing is incorporated through $\Delta_{ij}=\Delta_{ji}$. 
We then define a gauge invariant $d$-wave order parameter on each lattice site $i$
\begin{eqnarray}
 \Delta_i^d = \frac{1}{4} \left(  \Delta^d_{i,i+\hat{x}} +
       \Delta^d_{i,i-\hat{x}} - \Delta^d_{i,i+\hat{y}} - \Delta^d_{i,i-\hat{y}}\right),
\end{eqnarray}
where $\Delta^d_{i,j} = \Delta_{ij} \exp[-{\rm i}(\pi/\Phi_0) \int_{{\bf r}_j}^{{\bf r}_i} {\bf A(r)} \cdot d{\bf r}]$.
The chemical potential $\mu$ is adjusted to fix the average charge density $n = \frac{1}{N}\sum_i \langle n_i \rangle$,
while the electron number operator for spin $\sigma$ at site $i$ is given by
%\begin{eqnarray}
$n_{i\sigma} = c_{i\sigma}^{\dagger} c_{i\sigma}^{},  $
%n_i &=& n_{i\uparrow} + n_{i\downarrow}.
%\sigma_i^z &=& n_{i\uparrow} - n_{i\downarrow}. \label{eq:sd}
%\end{eqnarray}
%The average charge density at each site
and the local charge-density operator by 
%\begin{eqnarray}
$n_i =  n_{i\uparrow} + n_{i\downarrow},$
%\langle n_i \rangle &=& \langle n_{i\uparrow} \rangle + \langle n_{i\downarrow} \rangle
%\end{eqnarray}
respectively. $S_i^z = \frac{1}{2} \sigma_i^z = \frac{1}{2} \left( n_{i\uparrow}  - n_{i\downarrow} \right)$ 
is the $z$-component of the spin-operator at site $i$.
%, while$\sigma_i^z$ is defined by
%\begin{eqnarray}
%\langle \sigma_i^z \rangle &=& \langle n_{i\uparrow} \rangle -\langle  n_{i\downarrow} \rangle. %\label{eq:sd}
%$\sigma_i^z  = n_{i\uparrow}  - n_{i\downarrow}.$% \label{eq:sd}
%\end{eqnarray}

The Bogoliubov transformation
 \begin{eqnarray}
c_{i\sigma} = \sum_n \left(u_{in\sigma} \gamma_{n\sigma}^{} + v^{*}_{in\sigma} \gamma_{n-\sigma}^{\dagger} \right),
\label{eq:bogtrans}
\end{eqnarray}
diagonalizes the Hamiltonian in equation (\ref{eq:h}), which thereby takes the form
\begin{eqnarray}
 H = E_0 + \sum_{n\sigma} E_{n\sigma} \gamma_{n\sigma}^{\dagger} \gamma_{n\sigma}^{}.
 \label{eq:h_diag}
\end{eqnarray}
$E_0$ is the ground-state energy and $\gamma_{n\sigma}^\dag$  creats an elementary fermionic Bogoliubov 
quasiparticle excitation with quantum number $n$, spin $\sigma$, and energy $E_{n\sigma}>0$. 
%called a Bogoliubov quasiparticle.   
%, while $E_0$ is the ground state energy.
%The creation of an excitation leads to an increase of the total energy. %above the ground state energy.
%Thus the sum in equation (\ref{eq:h_diag}) describes the enhancement of energy above the ground state energy
%$E_0$. We stress that the summation in (\ref{eq:bogtrans}) and (\ref{eq:h_diag}) therefore runs only over $n$
%belonging to positive $E_{n\sigma}$.
%% because  the $E_{n\sigma}$ describ the exitation energies above the ground state energy $E_0$
%%We sum only over positive $E_{n\sigma}$, because
%%these denote denote the excitation energy, which increase the ground state energy $E_0$.
%%Calculating the commutators $[c_{i\sigma}, H]$
%With that we can derive the Bogoliubov-de Gennes equations. This is accomplished by determining the
%commutators $[c_{i\sigma}, H]$ for both spin directions $\sigma = \uparrow, \downarrow$ and
%performing the Bogoliubov transformation (\ref{eq:bogtrans}). We then end up with two equations, where
%we compare the coefficients of $\gamma_{n\sigma}$ and $\gamma_{n\sigma}^{\dagger}$. This leads us
%to the following two matrix equations
Calculation of the commutators of $H$ from equation (\ref{eq:h_diag}) with the electron operators $c_{i\sigma}$ leads 
to a Schr\"odinger-like set of BdG equations %called the Bogoliubov-de Gennes equations,
\begin{eqnarray}
\sum_j \begin{pmatrix} H_{ij}^{+} & \Delta_{ij} \\ \Delta_{ij}^* & -{H_{ij}^{-}}^* \end{pmatrix}
\begin{pmatrix} u_{jn\uparrow} \\ v_{jn\downarrow} \end{pmatrix} = E_{n\uparrow}
\begin{pmatrix} u_{in\uparrow} \\ v_{in\downarrow}\end{pmatrix},
\label{eq:bdg1}
\end{eqnarray}
and
\begin{eqnarray}
\sum_j \begin{pmatrix} H_{ij}^{+} & \Delta_{ij} \\ \Delta_{ij}^* & -{H_{ij}^{-}}^* \end{pmatrix}
\begin{pmatrix} v_{jn\uparrow}^* \\ u_{jn\downarrow}^* \end{pmatrix} = -E_{n\downarrow}
\begin{pmatrix} v_{in\uparrow}^* \\ u_{in\downarrow}^* \end{pmatrix},
\label{eq:bdg2}
\end{eqnarray}
with
\begin{eqnarray}
  H^{\pm}_{ij} = -t_{ij} + \delta_{ij} \left[-\mu+\frac{U}{2}
                 \left(\langle n_i \rangle \mp\langle \sigma_i^z \rangle \right) + V_i^{imp} \right].
\end{eqnarray}
%As mentioned above, we only
As we only search for solutions with positive $E_{n\sigma}$, it is sufficient to solve the
following single matrix equation
%we just have to solve the following equation
%he coefficients $u_n$ and $v_n$ have to fulfill the equation
\begin{eqnarray}
 \sum_j
  \begin{pmatrix}
  H^+_{ij} & \Delta_{ij} \\ \Delta_{ij}^{*} & -H_{ij}^{-*}
  \end{pmatrix}
  \begin{pmatrix}
   u_{jn}  \\ v_{jn}
  \end{pmatrix}
  = E_n
  \begin{pmatrix}
   u_{in}  \\ v_{in}
  \end{pmatrix},
\label{eq:bdg}
\end{eqnarray}
This is because the solutions for $E_n > 0$ are obviously identical to the solutions for the (positive) eigenvalues
$E_{n\uparrow}$ of equation (\ref{eq:bdg1})
\begin{eqnarray}
\hspace{-.3cm}
 E_n > 0: \qquad
\begin{pmatrix} u_{in\uparrow} \\ v_{in\downarrow} \end{pmatrix} =  \begin{pmatrix} u_{in}  \\ v_{in} \end{pmatrix}
 \text{and } E_{n\uparrow}  = E_n > 0,
\end{eqnarray}
while for $E_n < 0$ %the solutions of (\ref{eq:bdg}) are related to that of equation (\ref{eq:bdg2}) in the
%following way
the following relation holds
 %$-E_{n\downarrow}$ ($E_{n\downarrow} > 0$) of equation (\ref{eq:bdg2})
\begin{eqnarray}
 E_n < 0: \qquad
\begin{pmatrix} v_{in\uparrow}^* \\ u_{in\downarrow}^* \end{pmatrix} =  \begin{pmatrix} u_{in}  \\ v_{in} \end{pmatrix}
\text{and }  E_{n\downarrow} = - E_n > 0.
\end{eqnarray}
Since the solutions of equations (\ref{eq:bdg1}) and (\ref{eq:bdg2}) can be mapped on to those of the BdG 
equation (\ref{eq:bdg}), we diagonalize equation  (\ref{eq:bdg}) to obtain the pairing potential $\Delta_{ij}$, the charge density 
$\langle n_i \rangle$, and the local magnetization $\langle \sigma_i^z \rangle$  self-consistently from
\begin{eqnarray}
 \Delta_{ij} &=& \frac{V_d}{4} \sum_n \left(u_{in} v_{jn}^{*} + u_{jn} v_{in}^{*} \right)
               \tanh \left(\frac{E_n}{2 k_B T}\right),\\
 \langle n_{i\uparrow} \rangle  &=& \sum_n |u_{in}|^2 f(E_n), \\
 \langle n_{i\downarrow} \rangle  &=& \sum_n |v_{in}|^2 (1-f(E_n)),  \\
 \langle \sigma_i^z \rangle &=& \langle n_{i\uparrow} \rangle \: - \: \langle n_{i\downarrow} \rangle,
\end{eqnarray}
where $f(E_n)=(1+e^{E_n/k_BT})^{-1}$ is the Fermi distribution function and $T$ is the temperature. 
Sums over $n$ run over positive and negative  energies $E_n$. 

\begin{figure}[t!]
\centering
\vskip2mm
\includegraphics[width=7cm]{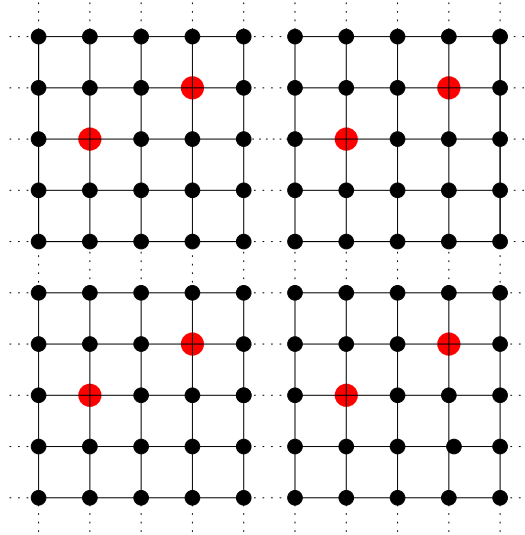}
\caption{\setlength{\baselineskip}{5mm}
Division of the lattice into identical supercells.  Lattice sites belonging to the same supercell are connected
via solid lines, while dashed lines link sites of different supercells. Red lattice sites simulate a possibly
existing disorder.}
\label{Suppfig1}
\end{figure}

\begin{figure}[t!]
\centering
\vskip2mm
\includegraphics[width=11cm]{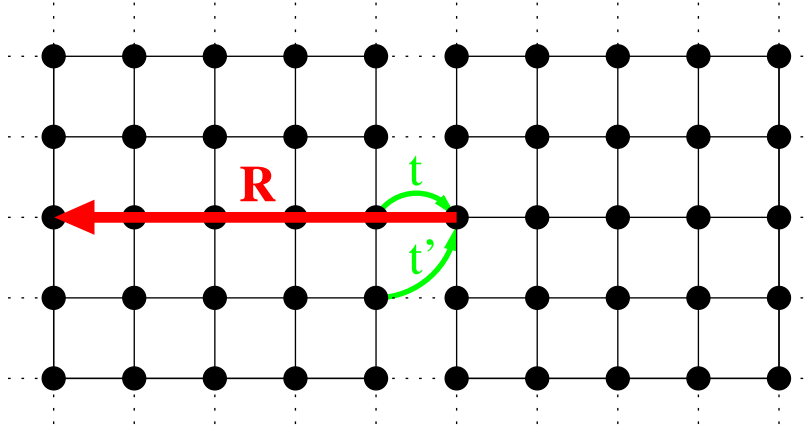}
\caption{\setlength{\baselineskip}{5mm}
%Nearest-neighbor hopping $t$ or next-nearest neighbor hopping $t'$ of a particle
Hopping between supercells.
%The translation vector ${\bf R}$ maps
%lattice sites of supercell 2 back into supercell 1.
A particle, which hops from the left supercell into the right supercell, is mapped back
to the left supercell through the translation vector ${\bf R}$. As a result the wave functions
$u$ and $v$ obtain an additional phase given by the magnetic Bloch theorem (\ref{eq:magnetic_bloch}).}
\label{Suppfig2}
\end{figure}

%Supercell method
%As we want to examine among others local density of states, we need to perform our calculations on
%lattice sizes as large as possible in order to obtain enough eigenvalues to get even at a
%high energy resolution smooth LDOS curves. The limiting factor for the system sizes
%is the computational time needed for diagonalizing the matrix in equation (\ref{eq:bdg}).
%Therefore 
To maximize the size of the system for which equation (\ref{eq:bdg}) can be diagonalized numerically, 
we take advantage of the magnetic translation 
symmetry of our model Hamiltonian (\ref{eq:h}) by dividing the lattice into $M_x \times M_y$ identical supercells
each with $N_x \times N_y$ sites (see Supp. Fig. {\bf\ref{Suppfig1}}).\cite{ghosal, zhuting, atkinson} 
%Hence we divide our lattice into $M_x \times M_y$
%identical supercells each with $N_x \times N_y$ sites.
We define the following magnetic translation operator \cite{Brown}
\begin{eqnarray}
 \mathcal T_ {\bf R} = \exp\left(-{\rm i} \;{\bf R} \cdot ({\bf k} + \frac{q}{c\hbar} {\bf A})\right),
 \label{transl:op}
\end{eqnarray}
where ${\bf R}$ is the translation vector and $\mathcal T_ {\bf R}$ translates
any lattice vector ${\bf r}$ to the position
${\bf r + R}$. Because $[H, \mathcal T_{\bf R}] = 0$, it is possible to block diagonalize
the Hamiltonian $H$ in equation (\ref{eq:h}) using the eigenstates of $ \mathcal T_{\bf R}$.
This reduces the eigenvalue problem (\ref{eq:bdg}) of dimension $2M_x N_x \times 2M_y N_y$ to
$M_x \times M_y$ eigenvalue equations of dimension $2N_x \times 2N_y$. Applying the
magnetic Bloch theorem
\begin{eqnarray}
 \begin{pmatrix} u_{n{\bf k}}(\mathcal T_{\bf R}{\bf r}_i) \\ v_{n{\bf k}}(\mathcal T_{\bf R}{\bf r}_i) \end{pmatrix} =
e^{-{\rm i} {\bf k \cdot R}}
\begin{pmatrix} e^{-{\rm i} \frac{\pi}{\Phi_0} {\bf A(R) \cdot r}_i} u_{n{\bf k}}({\bf r}_i)
\\ e^{{\rm i} \frac{\pi}{\Phi_0} {\bf A(R) \cdot r}_i} v_{n{\bf k}}({\bf r}_i) \end{pmatrix}
\label{eq:magnetic_bloch}
\end{eqnarray}
block diagonalizes the BdG equations (\ref{eq:bdg}), where ${\bf k} = 2\pi (\frac{n_x}{M_x N_x}$,
$\frac{n_y}{M_y N_y})$, $u_{n{\bf k}}({\bf r}_i) = u_{in{\bf k}}$ and $v_{n{\bf k}}({\bf r}_i) = v_{in{\bf k}}$.
Thus we have to solve the following $2N_x \times 2N_y$ matrix equation for each ${\bf k}$ value
\begin{eqnarray}
 \sum_j
  \begin{pmatrix}
   H^+_{ij}({\bf k}) & \Delta_{ij}({\bf k}) \\ \Delta_{ij}^{*}({\bf k}) & -H_{ij}^{-*}({\bf k})
  \end{pmatrix}
  \begin{pmatrix}
   u_{jn{\bf k}}  \\ v_{jn{\bf k}}
  \end{pmatrix}
  = E_{n{\bf k}}
  \begin{pmatrix}
   u_{in{\bf k}}  \\ v_{in{\bf k}}
  \end{pmatrix},
\end{eqnarray}
where
\begin{eqnarray}
%  H^{\pm}_{ij}({\bf k}) &=& -t_{ij}({\bf k}) + \delta_{ij} \left[-\mu+\frac{U}{2}
%                 \left(\langle n_i \rangle \mp\langle \sigma_i^z \rangle \right) + V_i^{imp} \right], \\
%\end{eqnarray}
%The pairing potential $\Delta_{ij}$, the average charge density $\langle n_i \rangle$ and the
%magnetization $\langle \sigma_i^z \rangle$ are calculated self-consistently from
%\begin{eqnarray}
 \Delta_{ij} &=& \frac{V_d}{4 M_x M_y} \sum_{n{\bf k}} \left(u_{in{\bf k}} v_{jn{\bf k}}^{*}
                 + u_{jn{\bf k}} v_{in{\bf k}}^{*} \right)
               \tanh \left(\frac{E_{n{\bf k}}}{2 k_B T}\right),\\
 \langle n_{i\uparrow} \rangle  &=& \frac{1}{M_x M_y} \sum_{n{\bf k}} |u_{in{\bf k}}|^2 f(E_{n{\bf k}}), \\
 \langle n_{i\downarrow} \rangle  &=& \frac{1}{M_x M_y} \sum_{n{\bf k}} |v_{in{\bf k}}|^2 (1-f(E_{n{\bf k}})). % \\
% \langle \sigma_i^z \rangle &=& \langle n_{i\uparrow} \rangle \: - \: \langle n_{i\downarrow} \rangle,
\end{eqnarray}
$H_{ij}$ and $\Delta_{ij}$ are only ${\bf k}$ dependent, if $i$ and $j$ belong to different supercells.
Then the back-mapping (see Supp. Fig. {\bf\ref{Suppfig2}}) leads to an additional phase for the $u$'s and $v$'s
according to (\ref{eq:magnetic_bloch}), which is assigned to the matrix elements $t_{ij}(\bf k)$ and
$\Delta_{ij}(\bf k)$. To make sure that two magnetic translations commute, we have to choose the
magnetic field such that its flux through every supercell is a multiple of $2\Phi_0$. \cite{zhuting, Brown} 
Hence $2\Phi_0$ provides a lower boundary for the magnetic flux threading each supercell, which 
corresponds for a supercell enclosing an area of e.g. $22a \times 22a$ to a magnetic field of about 59 T 
% given that each supercell has an area of $22a \times 22a$ 
(we assumed a typical value for the in-plane lattice constant $a$ in the cuprates of about $a=3.8$\AA).  

%We constructed a lattice consisting consisting of identical supercells. Arriving at the edges of our
%superlattice we have to make sure that the Born-von Karman boundary condition
%\begin{eqnarray}
% \mathcal T (M_i N_i{\bf e}_i) \Psi_i = \Psi_i,
%end{eqnarray}
%is valid. $N_i$ is the length of the supercell (the lattice constant $a$ is set to unity) and $M_i$ is the number of %supercells in the ${\bf e}_i$ direction and $\Psi_i$ is a placeholder for the $u_i$'s and $v_i$'s. It then follows
%\begin{eqnarray}
% M_i N_i {\bf e}_i \cdot ({\bf k} + \frac{q}{c\hbar} {\bf A}) &=&
%M_i N_i \left(\frac{2 \pi n_i}{M_i N_i}+\frac{\pi}{\Phi_0} A_i \right) \\
%&=&  2 \pi n_i + \frac{\pi}{\Phi_0} A_i  M_i N_i = 2 \pi \tilde{l},
%\end{eqnarray}
%while $\tilde{l}$ is integer. To fulfill this constraint the second addend must
%be a multiple of $2 \pi$:
%\begin{eqnarray}
% M_y N_y \frac{\pi}{\Phi_0} H x = M_y N_y \frac{\pi}{\Phi_0} \frac{\Phi}{N_x N_y} x
%= 2 \pi l
%\end{eqnarray}
%Because this must hold for every $x$, we end up with
%\begin{eqnarray}
%\frac{M_y}{N_x} \frac{\Phi}{\Phi_0} = 2 l.
%\end{eqnarray}

%In order to make comparison with experimental results we determine the local density of states (LDOS)
%To compare our results with existing STM experiments we determine the local density of states (LDOS)
%\begin{eqnarray}
% N_i = \frac{1}{M} \sum_{n {\bf k}} \left(|u_{in{\bf k}}|^2 \delta(E - E_{n{\bf k}})
%       + |v_{in{\bf k}}|^2 \delta(E + E_{n{\bf k}})\right),
%\end{eqnarray}
%which is directly proportional to the differential conductance. 

In order to make contact with neutron scattering
experiments, we evaluate the magnetic structure factor $S({\bf q})$. %, which we approximate as follows.
In homogeneous systems it is defined as
\begin{eqnarray}
 S({\bf q}) = \frac{1}{N} \sum_i \langle \sigma_i^z \sigma_0^z \rangle  e^{-{\rm i} {\bf q} \cdot ({\bf r}_i -{\bf r}_0)}.
\end{eqnarray}
We approximate the spin-spin correlation function by the following factorization
\begin{eqnarray}
\langle \sigma_i^z \sigma_0^z \rangle  \rightarrow \langle \sigma_i^z \rangle \langle \sigma_0^z \rangle.
 \end{eqnarray}
Because the system which we are interested in is in general inhomogeneous, we have to sum over all lattice
sites. Hence we find the expression
\begin{eqnarray}
|M({\bf q})|^2 = \frac{1}{N^2} \sum_{ij} \langle \sigma_i^z \rangle \langle  \sigma_j^z \rangle e^{-{\rm i} {\bf q} \cdot ({\bf r}_j - {\bf r}_i)}.
 \end{eqnarray}
This approximation of the magnetic structure factor is identical to the Fourier transform of the magnetization squared.
%The LDOS is directly proportional to the differential conductance measured by
%scanning tunneling microscopy (STM), while the magnetic structure factor
%
%Moreover we investigate the staggered magnetization at each lattice site $i$
%\begin{eqnarray}
% M_i^s = (-1)^i \langle \sigma_i^z \rangle
%\end{eqnarray}
%to identify phase boundaries of the antiferromagnetic order more easily.

%{\bf PH--I would end Supplementary Info here.  I have not edited what follows.}

\begin{figure}[p]

{\large $\Phi = 0$ \hspace{6.6cm} $\Phi = 2\Phi_0$}

\hspace{-2mm}
\begin{overpic}[scale=1.025,unit=1mm]{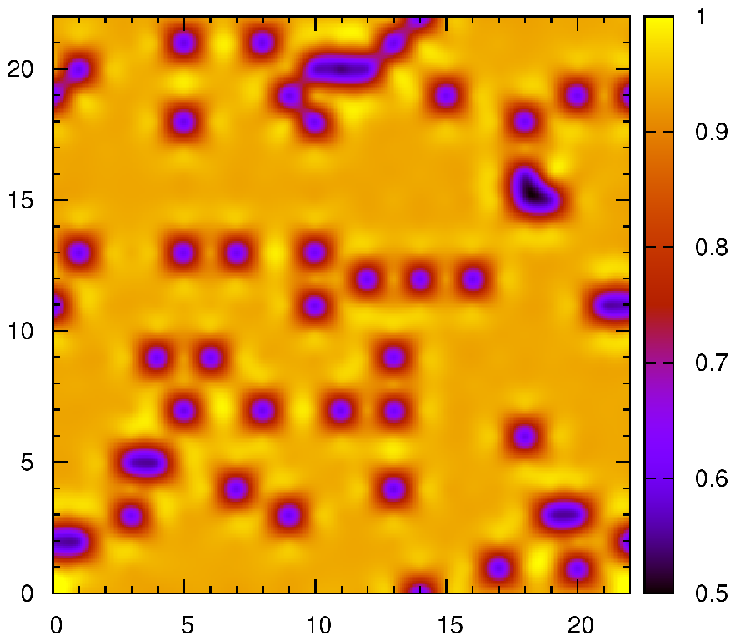}
    \put(-3,61){{\bf a}}
 \end{overpic}
\hspace{.35cm}
\begin{overpic}[scale=1.025,unit=1mm]{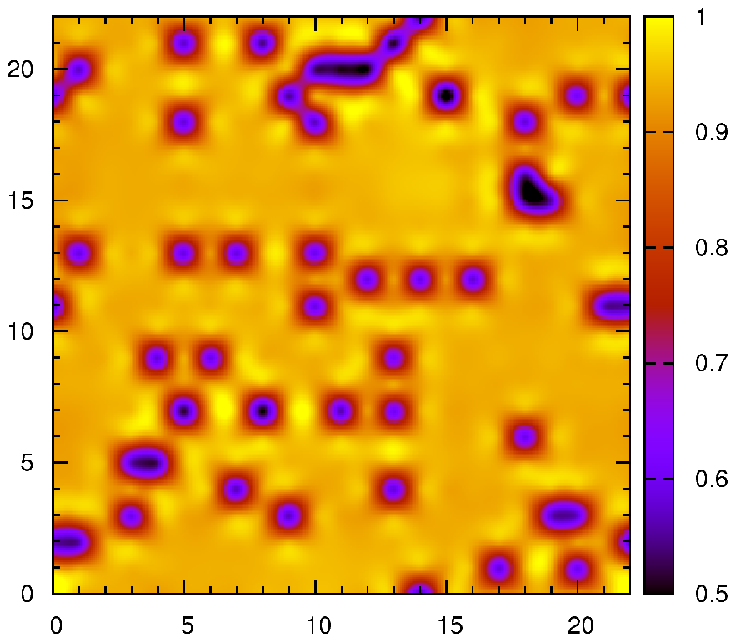}
    \put(-3,61){{\bf d}}
 \end{overpic}

 \begin{overpic}[scale=1.03,unit=1mm]{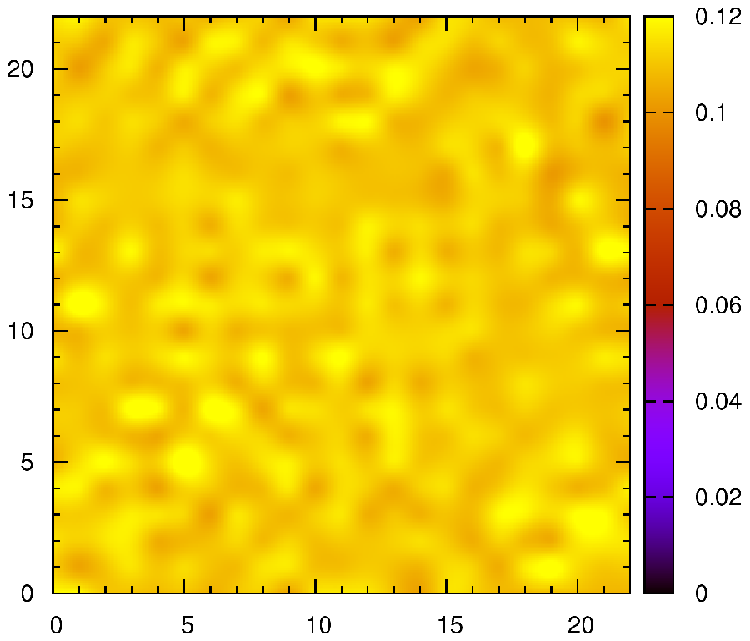}
    \put(-3,61){{\bf b}}
 \end{overpic}
\hspace{0.35cm}
 \begin{overpic}[scale=1.03,unit=1mm]{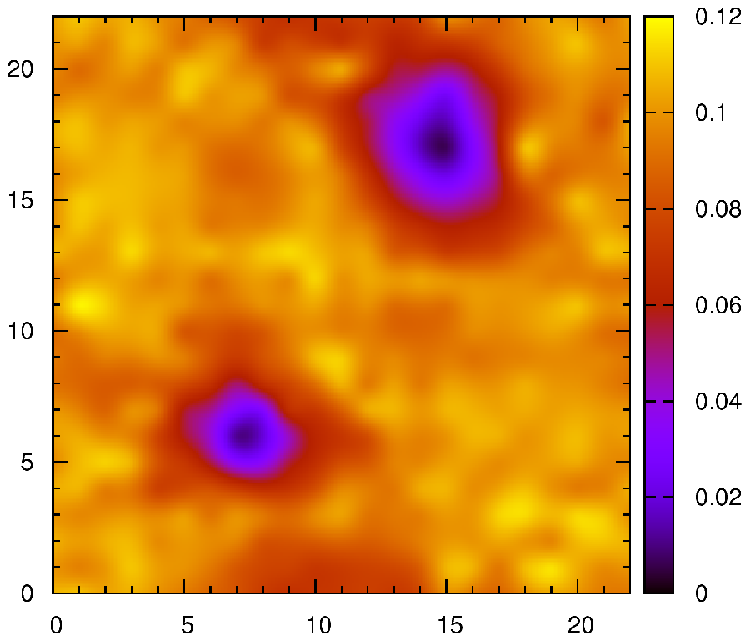}
    \put(-3,61){{\bf e}}
 \end{overpic}

 \begin{overpic}[scale=1.03,unit=1mm]{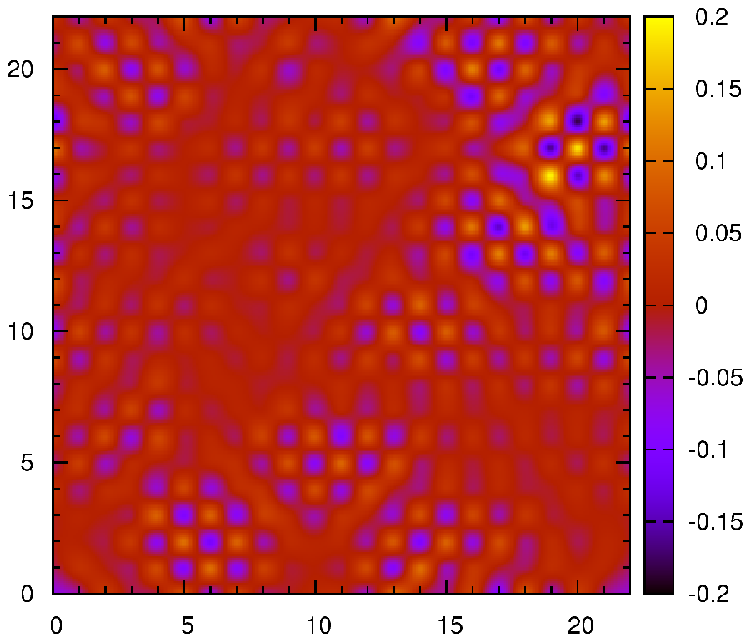}
    \put(-3,61){{\bf c}}
 \end{overpic}
\hspace{0.35cm}
 \begin{overpic}[scale=1.03,unit=1mm]{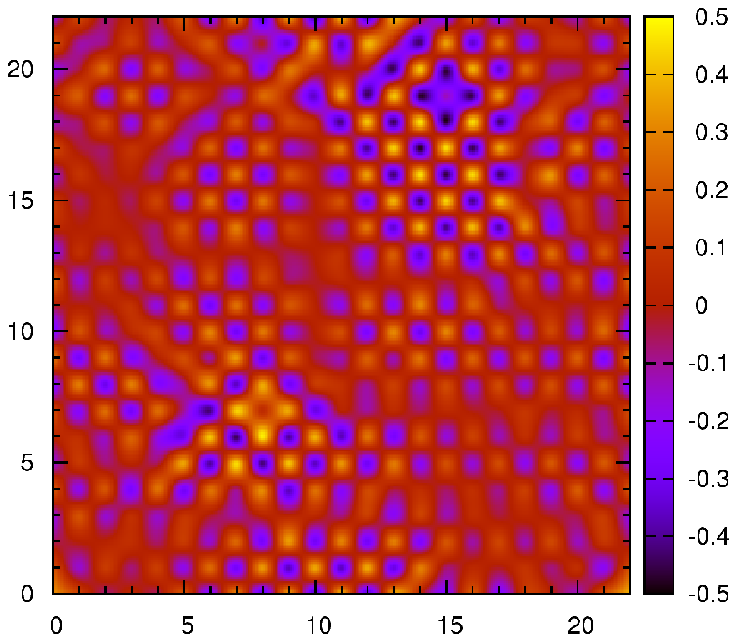}
    \put(-3,61){{\bf f}}
 \end{overpic}
\caption{\setlength{\baselineskip}{5mm}
 {\bf Switching on an orbital magnetic field.} {\bf a-c}. Zero-field data. {\bf d-f}. Finite-field data ($\Phi = 2 \Phi_0$).
{\bf a, d} show the charge density $\langle n_i \rangle$, {\bf b, e} the $d$-wave order parameter $\Delta_i^d$, and {\bf c, f} the 
magnetization $\langle \sigma_i^z \rangle$ in real space. The same set of parameters is used here as in the rest of
the paper, i.e. $x = 10\% = n_{imp}, U=2.9t, V_{imp} = 1.3t$. These data were obtained at the lowest temperature 
$T=0.025t$ we considered throughout the paper. Note the different scales in {\bf c} and {\bf f}.}
\label{Suppfig3}
\end{figure}

In Supp. Fig. \ref{Suppfig3} results for $\langle n_i \rangle$, $\Delta_i^d$, and $\langle \sigma_i^z \rangle$ are
 shown in zero field (left column) and in finite magnetic
field (right column) for a typical impurity configuration. One can identify the location of the impurities by the 
point-like suppression of the charge density (top row). While the $d$-wave
order parameter is nearly homogeneous in the zero-field case (see Supp. Fig. {\bf \ref{Suppfig3}b}), one can clearly spot the
positions of the two vortices where $\Delta_i^d$ is
suppressed to zero in Supp. Fig. {\bf \ref{Suppfig3}e}. In comparison to the zero-field case, a finite orbital magnetic field leads to an additional 
reduction of the 
order parameter over the entire lattice. Finally, for the parameters used here, the zero
field case already contains impurity-induced antiferromagnetic order (see Supp. Fig. {\bf \ref{Suppfig3}c}), which is significantly enhanced by
switching on a magnetic field. The magnetization peaks near the vortex cores, but due to the fact that strong type-II superconductors 
are penetrated by the field much beyond the cores, the magnetization is also enhanced in regions far away from the vortices, 
where the order parameter is nearly homogeneous. The SDW emerges due to the splitting of 
the Andreev bound state as explained in greater detail in the main body of the paper.

\vspace{5mm}
\noindent
\textbf{References}
{}
%\vspace*{-1cm}

\vspace{5mm}
\noindent
{\bf Acknowledgements}

\noindent This work was supported by the DFG through SFB 484 (M.S. and
A.P.K.), by The Danish Council for Independent Research $|$
Natural Sciences (B.M.A.), and by the DOE under grant
DE-FG02-05ER46236 (P.J.H.).

\vspace{5mm}

%\textbf{Figure Legends}\\

\end{document}